\documentclass[utf8]{frontiersinFPHY_FAMS} 

\setcitestyle{square} 
\usepackage{url,hyperref,microtype,subcaption}
\usepackage[onehalfspacing]{setspace}

\def\keyFont{\fontsize{8}{11}\helveticabold }
\def\firstAuthorLast{Jose Luis Bl\'azquez-Salcedo {et~al.}} 
\def\Authors{Jose Luis Bl\'azquez-Salcedo\,$^{1,*}$, Luis Manuel Gonz\'alez-Romero\,$^{1}$, \\ Fech Scen Khoo\,$^{2}$, Jutta Kunz\,$^{2}$, and Vincent Preut\,$^{2}$}
\def\Address{$^{1}$Departamento de F\'isica Te\'orica and IPARCOS, Facultad de Ciencias F\'isicas, 
	Universidad Complutense de Madrid, Spain \\
$^{2}$Institute of  Physics, University of Oldenburg, Postfach 2503, D-26111 Oldenburg, Germany }
\def\corrAuthor{Jose Luis Bl\'azquez-Salcedo}

\def\corrEmail{jlblaz01@ucm.es}

\begin{document}
\onecolumn
\firstpage{1}

\title[$\phi$-modes of neutron stars in a massless STT]{$\phi$-modes of neutron stars in a massless scalar-tensor theory} 

\author[\firstAuthorLast ]{\Authors} 
\address{} 
\correspondance{} 

\extraAuth{Luis Manuel Gonz\'alez-Romero \\ mgromero@ucm.es \\
Fech Scen Khoo \\  fech.scen.khoo@uni-oldenburg.de \\ Jutta Kunz \\  jutta.kunz@uni-oldenburg.de \\ Vincent Preut \\  vincent.preut@uni-oldenburg.de}

\maketitle

\begin{abstract}

\section{}
Scalar-tensor theories allow for a rich spectrum of quasinormal modes of neutron stars. 
The presence of the scalar field allows for polar monopole and dipole radiation, as well as for additional higher multipole 
modes
led by the scalar field. 
Here we present these scalar-led $\phi$-modes for the lowest multipoles, $l=0$, 1 and 2 for a massless scalar-tensor theory of the Brans-Dicke type, motivated by $R^2$ theory, and compare with those of a minimally coupled scalar field in general relativity.
We consider a set of six realistic equations of state and extract universal relations for the modes.

\tiny
 \keyFont{ \section{Keywords:} Neutron stars, gravitational waves, quasinormal modes, scalar-tensor theory, equation of state, universal relations} 
\end{abstract}

\section{Introduction}

Currently astrophysical observations are well described by Einstein's general relativity, not only in the solar system but also in the realm of strong gravity, where neutron stars and black holes reside \citep{Will:2018bme}.
Still, the quest for a more fundamental theory of gravity compatible with quantum mechanics as well as the enigmatic presence of the dark sector in cosmology are motivating studies of alternative theories of gravity  \citep{Saridakis:2021vue}.
Compact objects like black holes and neutron stars then constitute excellent astrophysical laboratories to provide constraints on such alternative theories of gravity \citep{Faraoni:2010pgm,Berti:2015itd,Saridakis:2021vue,LISA:2022kgy}.

With the advent of gravitational wave observations here a new window to the Universe was opened \citep{LIGOScientific:2016aoc,LIGOScientific:2017vwq,LIGOScientific:2017ync,LIGOScientific:2020aai,LIGOScientific:2021qlt}.
In particular, in connection with multi-messenger astronomy new insights on the properties of neutron stars have been gained.
Like the collision of black holes, the collision of neutron stars is governed by three phases, inspiral, merger and ringdown.
The gravitational radiation emitted during the ringdown is associated with the quasinormal modes of the newly created compact object \citep{Andersson:1996pn,Andersson:1997rn,Kokkotas:1999bd}.
Therefore investigation of the spectrum of quasinormal modes of black holes and neutron stars represents an important step for our understanding.

For neutron stars a crucial open question is the composition of their inner core. 
Here numerous equations of state have been put forward since experiments on Earth cannot reach the extreme conditions of matter present in the interior of these stars \citep{Lattimer:2021emm}.
While observations of high mass pulsars have put strong constraints on the maximum mass of a neutron star obtained by a given equation of state \citep{Demorest:2010bx,Antoniadis:2013pzd,NANOGrav:2019jur}, additional constraints are due to the observation and analysis of merger events \citep{LIGOScientific:2017ync}.

While current gravitational wave observations do not yet constrain the quasinormal modes of neutron stars, future gravitational wave observatories should be sufficiently sensitive to do so in combination with further observations of neutron star properties \citep{Berti:2015itd}.
However, here another all-important property of neutron stars will enter, namely their universal relations  \citep{Yagi:2016bkt,Doneva:2017jop}.
Such universal relations represent relations between properly scaled neutron star properties, that show a large degree of independence of the equation of state.
They exist not only for intrinsic neutron star properties like their moment of inertia, quadrupole moment or Love number, but also for their quasinormal modes, where they were noticed already early on  \citep{Andersson:1996pn,Andersson:1997rn,Benhar:1998au}.
By now many studies of universal relations of quasinormal modes of neutron star in general relativity have been done (e.g.,
\citep{Andersson:1996pn,Andersson:1997rn,Benhar:1998au,Benhar:2004xg,Tsui:2004qd,Lau:2009bu,Blazquez-Salcedo:2012hdg,Blazquez-Salcedo:2013jka,Chirenti:2015dda,Lioutas:2021jbl,Sotani:2021nlx,Sotani:2021kiw,Zhao:2022tcw}).

Universal relations for quasinormal modes of neutron stars are not only of relevance with respect to our current lack of knowledge of the proper equation of state, they are also valuable to learn about alternative theories of gravity whenever these relations differ distinctly from those of general relativity \citep{Berti:2015itd,Doneva:2017jop}.
Axial quasinormal modes and their universal relations have already been considered for a variety of alternative theories of gravity \citep{Blazquez-Salcedo:2015ets,Blazquez-Salcedo:2018tyn,Blazquez-Salcedo:2018qyy,AltahaMotahar:2018djk,Blazquez-Salcedo:2018pxo,AltahaMotahar:2019ekm}.
Axial modes do not involve perturbations of the neutron star fluid and neither of the scalar field when, for instance, scalar-tensor theories of gravity are considered \citep{Brans:1961sx,Damour:1992we,Fujii:2003pa}. They are pure gravitational modes.

Polar quasinormal modes, on the other hand, involve besides the metric also the fluid and the scalar field, when present. Therefore early studies of polar modes in such alternative gravity theories have employed the Cowling approximation, where only fluid perturbations are taken into account \citep{Sotani:2004rq,Staykov:2015cfa}.
However, by now also the full set of perturbations has been considered when determining polar quasinormal modes of several alternative theories of gravity with a scalar degree of freedom \citep{Sotani:2014tua,Mendes:2018qwo,Blazquez-Salcedo:2020ibb,Kruger:2021yay,Dima:2021pwx,Blazquez-Salcedo:2021exm,Blazquez-Salcedo:2022pwc,Blazquez-Salcedo:2022dxh}.

Recently we have determined polar modes for a family of quadratic gravity theories where the Einstein-Hilbert action is supplemented by an $R^2$ term.
This type of action represents an interesting class of $f(R)$ theories \citep{Sotiriou:2008rp,DeFelice:2010aj,Capozziello:2011et}.
Transformation yields a Brans-Dicke type scalar-tensor theory with a particular coupling function and potential for the scalar field, where the coupling constant of $R^2$ theory determines the mass of the scalar field \citep{Yazadjiev:2014cza,Staykov:2014mwa}.
In the limit of infinite scalar mass general relativity with a minimally coupled scalar field is recovered, whereas in the limit of vanishing scalar mass a massless Brans-Dicke type scalar-tensor theory is obtained.

While we have previously determined polar quasinormal modes of neutron stars for this family of alternative gravity theories with finite values of the scalar mass \citep{Blazquez-Salcedo:2020ibb,Blazquez-Salcedo:2021exm,Blazquez-Salcedo:2022pwc,Blazquez-Salcedo:2022dxh}, we here focus on these two limiting cases and construct their scalar-led modes, i.e., their $\phi$-modes.
$\phi$ modes are present in the radial ($l=0$) and dipole ($l=1$) case besides the previously studied fluid F- and pressure H$_n$-modes.
They also arise in the quadrupole ($l=2$) case (as well as for $l>2$), where the fundamental fluid $f$ {mode} has previously been investigated.
We then determine universal relations for these $\phi$-modes and compare to general relativity.

The paper is structured as follows: In section 2 we present the theoretical setting, the massless Brans-Dicke type scalar-tensor theory, the equations for the background neutron star solution, as well as the metric, fluid and scalar perturbations. Subsequently we present our results on the $\phi$-modes and their universal relations, the quadrupole modes in section 3, the dipole modes in section 4 and the radial modes in section 5. We conclude in section 6. In the appendices we present tables for the universal relations.

\section{Theoretical Setting}

\subsection{Massless Brans-Dicke type scalar-tensor theory}

We here consider neutron stars that are governed in the Einstein frame by the action 
($G=c=1$) 
\begin{equation}
S [g_{\mu\nu},\phi] = \frac{1}{16\pi } \int d^4x \sqrt{-g}
\big( R - 2\partial_{\mu}\phi \, \partial^{\mu}\phi 
+ L_{M}(A^2(\phi)g_{\mu\nu},\chi) \big)~,
\label{EinsteinAction}
\end{equation}
where $R$ is the curvature scalar, $\phi$ denotes the massless scalar field, and we employ the Brans-Dicke coupling function
\begin{equation}
A(\phi)= e^{-\frac{1}{\sqrt{3}}\phi} 
\end{equation}
in the matter action $L_M$.
This action is obtained in the massless limit of $R^2$ theory, when transformed to a scalar-tensor theory in the Einstein frame, and in general features the potential term \citep{Yazadjiev:2014cza,Staykov:2014mwa}
\begin{equation}
V=\frac{3m_{\phi}^2}{2} \big(1- e^{-\frac{2\phi}{\sqrt{3}}}\big)^2 \ .
\label{pots}
\end{equation}
The presently considered massless theory and general relativity represent limiting cases for this type of theory.

Variation of the action (\ref{EinsteinAction}) leads to the Einstein equations 
\begin{equation}
G_{\mu\nu} = T^{(S)}_{\mu\nu} + 8 \pi T^{(M)}_{\mu\nu}
~,
\label{eq_G}
\end{equation}
with Einstein tensor $G_{\mu\nu} = R_{\mu\nu} - \frac{1}{2}Rg_{\mu\nu}$,
stress-energy tensor for the scalar field
\begin{equation}
T^{(S)}_{\mu\nu}=2\partial_{\mu}\phi\partial_{\nu}\phi -
g_{\mu\nu} \partial^{\sigma}\phi\partial_{\sigma}\phi
\ ,
\end{equation}
and stress-energy tensor for the neutron star matter
\begin{equation}
T^{(M)}_{\mu\nu} = (\rho + p)u_{\mu}u_{\nu} + pg_{\mu\nu}
~,
\label{T_matter}
\end{equation}
with pressure {$p$} and density {$\rho$} defined in terms of the physical pressure $\hat{p}$ and physical density $\hat{\rho}$ via the coupling function $A$,
\begin{equation}
p = A^4 \hat{p}~, \quad \rho = A^4 \hat{\rho} ~,
\label{p_rho_EJ}
\end{equation}
and with equation of state $\hat{p}(\hat{\rho})$.
Variation of the action with respect to the scalar field leads to the scalar field equation
\begin{equation}
\nabla_{\mu}\nabla^{\mu}\phi = -4\pi\frac{1}{A}\frac{dA}{d\phi} T^{(M)} 
~.
\label{scalar_eom}
\end{equation}

\subsection{Neutron star properties}

For the background metric of the neutron star we employ a static and spherically symmetric line element
\begin{equation}
ds^2 = g_{\mu\nu}^{(0)} dx^{\mu} dx^{\nu} = -e^{2\nu(r)} dt^2
+ e^{2\lambda(r)} dr^2 + r^2 (d\theta^2 + 
\text{sin}^2 \theta \, d\varphi^2 )
~.
\end{equation}
Accordingly, the scalar field and the pressure and the energy density of the fluid are parameterized by
\begin{eqnarray}
\phi=\phi_0(r) ~,  \hat{p}=\hat{p}_0(r) ~, \hat{\rho}=\hat{\rho}_0(r) ~,
\end{eqnarray}
while the fluid four-velocity
is given by
\begin{eqnarray}
u^{(0)}=-e^{\nu}dt ~.
\end{eqnarray}
The superscript $(0)$ and the subscript $0$ will always be used to indicate the background quantities.

Insertion of the ansatz into the set of field equations leads to the following set of background equations inside the star
\begin{eqnarray}
&&\frac{1}{r^2}\frac{d}{dr}\left[r(1- e^{-2\lambda})\right]= 8\pi
A_0^4 {\hat{\rho}_0} + e^{-2\lambda}\left(\frac{d\phi_0}{dr}\right)^2
~,  \label{eq_lambda} \\
&&\frac{2}{r}e^{-2\lambda} \frac{d\nu}{dr} - \frac{1}{r^2}(1-
e^{-2\lambda})= 8\pi A_0^4 \hat{p}_0 +
e^{-2\lambda}\left(\frac{d\phi_0}{dr}\right)^2 
V_0 
~, \label{eq_nu}
\\
&&\frac{d\hat{p}_0}{dr}= - (\hat{\rho}_0 + \hat{p}_0) \left(\frac{d\nu}{dr} +\frac{1}{A_0}\frac{dA_0}{d\phi_0}\frac{d\phi_0}{dr} \right) ~, \label{eq_pJ} 
\\
&&\frac{d^2\phi_0}{dr^2} + \left(\frac{d\nu}{dr} -
\frac{d\lambda}{dr} + \frac{2}{r} \right)\frac{d\phi_0}{dr}= 4\pi\frac{1}{A_0}\frac{dA_0}{d\phi_0} A_0^4(\hat{\rho}_0-3\hat{p}_0)e^{2\lambda} 
~, \label{eq_phi} 
\end{eqnarray}
where $A_0=A(\phi_0)$. 

In order to obtain the moment of inertia in first order perturbation theory for slow rotation, we now introduce the angular velocity of the star $\Omega$, which enters the line element via the inertial dragging $\omega(r)=\Omega-w(r)$ \citep{Hartle:1967he,Sotani:2012eb},
\begin{equation}
ds^2 = -e^{2\nu(r)} dt^2
+ e^{2\lambda(r)} dr^2 + r^2 (d\theta^2 + 
\text{sin}^2 \theta \, d\varphi^2 )-2(\Omega-w)r^2 \text{sin}^2 \theta \, dt \, d\varphi
~.
\end{equation}
The fluid four-velocity changes to 
\begin{eqnarray}
u=-e^{\nu}(dt+r^2\sin^2{\theta} \, w \, d\varphi) ~,
\end{eqnarray}
while the pressure and the energy density do not change to lowest order.
The resulting equation for $w$ is then
\begin{equation}
\frac{e^{\nu-\lambda}}{r^4}\frac{d}{dr}\left[ e^{-(\nu+\lambda)}r^4\frac{dw}{dr}\right]= 16\pi
A_0^4 {(\hat{\rho}_0+\hat{p}_0)}w \ .
\end{equation}

From the asymptotic behaviour of the metric functions the mass $M$ and the angular momentum $J$ of the star are obtained
\begin{eqnarray}
e^{2\nu}=e^{-2\lambda} &\sim& 1-2M/r ~, \\
w &\sim& \frac{2J}{r^3} ~, 
\end{eqnarray}
and the moment of inertia is given by $I=J/\Omega$.
The massless scalar field is long-ranged
\begin{eqnarray}
\phi_0 &\sim& \frac{1}{r}~. 
\end{eqnarray}

We require the star to be regular at the center and thus possesses finite central values of the pressure, the density and the scalar field
\begin{eqnarray}
\nu(0) = \nu_c ~, \ \ \lambda(0) = 0 ~, \ \ \phi_0(0) = \phi_c ~, \\
\hat{p}(0)=\hat{p}_c ~, \ \ \hat{\rho}(0)=\hat{\rho}_c ~, \\
w(0) =0 ~.
\end{eqnarray}
The boundary of the star $r=R$ is determined by the condition of vanishing pressure $\hat{p}(R)=0$. 

\subsection{Polar perturbations}

Dominated by the scalar field, the $\phi$-modes are polar perturbations.
To study such $\phi$-modes we therefore briefly recall the polar perturbations for neutron stars in scalar-tensor theories.
(For further details on the derivation see e.g. \citep{Regge:1957td,Zerilli:1970se,Thorne:1967a,Price:1969,Thorne:1969rba,Campolattaro:1970,Thorne:1980ru,Detweiler:1985zz,Chandrasekhar:1991fi,Chandrasekhar:1991,Chandrasekhar:1991_,Ipser:1991ind,Kojima:1992ie}.)

The static and spherically symmetric solutions constitute the zeroth order background functions.
We now introduce the perturbation parameter $\epsilon \ll 1$, in order to keep track of the order of the perturbations, when we perturb the background metric, the neutron star fluid, and the scalar field.
The general set of perturbations reads
\begin{eqnarray}
g_{\mu\nu} = g_{\mu\nu}^{(0)}(r) + \epsilon h_{\mu\nu}(t,r,\theta,\varphi)~, \\
p = p_0(r) + \epsilon \delta p(t,r,\theta,\varphi)~, \\
\rho = \rho_0 (r) + \epsilon \delta\rho(t,r,\theta,\varphi)~,  \\
u_{\mu} = u^{(0)}_{\mu} (r) + \epsilon \delta u_{\mu} (t,r,\theta,\varphi)
~, \\
\label{Phiperturb}
\phi = \phi_{0}(r) + \epsilon \delta\phi (t,r,\theta,\varphi)~.
\end{eqnarray}

Focusing now on polar perturbations, the metric perturbations are specified as
\begin{equation}
h_{\mu\nu}^{(\text{polar})} = \sum\limits_{l,m}\,\int    
\left[
\begin{array}{c c c c}
r^l e^{2\nu} H_0 Y_{lm} & -i \omega r^{l+1} H_1 Y_{lm} & 
0 & 0 \\
-i \omega r^{l+1} H_1 Y_{lm} &  r^l e^{2\lambda} H_2 Y_{lm} & 
0 & 0 \\
0 & 0
& r^{l+2} K  Y_{lm} & 0
\\
0 & 0  & 
0 & r^{l+2}  \sin^2\theta K  Y_{lm} \\
\end{array}
\right]
e^{-i\omega t} d\omega
~,
\end{equation}
with the spherical harmonics $Y_{lm}$, where $l$ and $m$ denote the multipolar indices (and $m$ will be set to zero in accordance with the spherically symmetric background).
The fluid perturbations read
\begin{equation}
\delta p = \sum\limits_{l,m}\,
\int   r^l \Pi_{1}  Y_{lm} e^{-i\omega t} d\omega~, \quad  
\delta \rho = \sum\limits_{l,m}\,
\int   r^l E_{1 } Y_{lm} e^{-i\omega t }d\omega~, \quad
\end{equation}
\begin{eqnarray}
\delta u_{\mu} = 
\sum\limits_{l,m}\,\int    
\left[
\begin{array}{c}
\frac{1}{2} r^l e^{\nu} H_0 Y_{lm}  \\
r^l i\omega e^{-\nu} 
\left(e^{\lambda}W/r -r H_1 \right) Y_{lm}  \\
-i\omega r^l e^{-\nu} V \partial_{\theta} Y_{lm}
\\
-i\omega r^l e^{-\nu} V \partial_{\varphi} Y_{lm}  \\
\end{array}
\right]
e^{-i\omega t} d\omega
~,
\end{eqnarray}
and the perturbation of the scalar field is given by 
\begin{equation} 
\delta \phi =  \sum\limits_{l,m}\,\int  r^l \phi_1 \, 
Y_{lm} e^{-i\omega t} d\omega ~.
\end{equation}
The perturbations have been decomposed not only with respect to the scalar, vector and tensor spherical harmonics, but also with respect to the complex frequency $\omega=\omega_R+i\omega_I$.
Here the real part $\omega_R$ corresponds to the characteristic frequency of the quasinormal modes, and the imaginary part $\omega_I$ represents their decay rate. 
Outside the star the perturbations simplify substantially, since the fluid is confined to the star, i.e., $\Pi_1=E_1=0$.

Substitution of this ansatz into the general set of field equations leads to a set of ordinary differential equations to be solved subject to adequate boundary conditions (see e.g., \citep{Blazquez-Salcedo:2020ibb,Blazquez-Salcedo:2021exm,Blazquez-Salcedo:2022pwc}).
In particular, to obtain the quasinormal modes we follow the same procedure as before \citep{Blazquez-Salcedo:2020ibb,Blazquez-Salcedo:2021exm,Blazquez-Salcedo:2022pwc}: 
for $r\to\infty$ we impose outgoing-wave conditions, and for the center $r=0$ and the surface $r=R$ of the star we require that the perturbation functions are regular. 
These physically motivated conditions then yield the necessary set of boundary conditions for the perturbation functions, that are implemented in the numerical procedure (see e.g., \citep{Blazquez-Salcedo:2018pxo} for more details on the employed numerical method).

In the following we will discuss the results of our calculations and analysis of the $l=0$ (radial or monopole) $\phi$-modes, the $l=1$ (dipole) $\phi$-modes and the $l=2$ (quadrupole) $\phi$-modes
for a set of six realistic equations of state describing
\begin{itemize}
	\itemsep=-1pt
	\item plain nuclear matter: SLy \citep{Douchin:2001sv} and APR4 \citep{Akmal:1998cf},
	\item a nucleon-hyperon fluid: GNH3 \citep{Glendenning:1984jr}, H4 \citep{Lackey:2005tk},
	\item hybrid nuclear-quark matter: ALF2 \citep{Alford:2004pf}, WSPHS3 \citep{Weissenborn:2011qu}.
\end{itemize}

\section{Quadrupole $\phi$-modes}

\subsection{Spectrum}

\begin{figure}[h!]
	\centering
	\includegraphics[width=.32\textwidth, angle =-90]{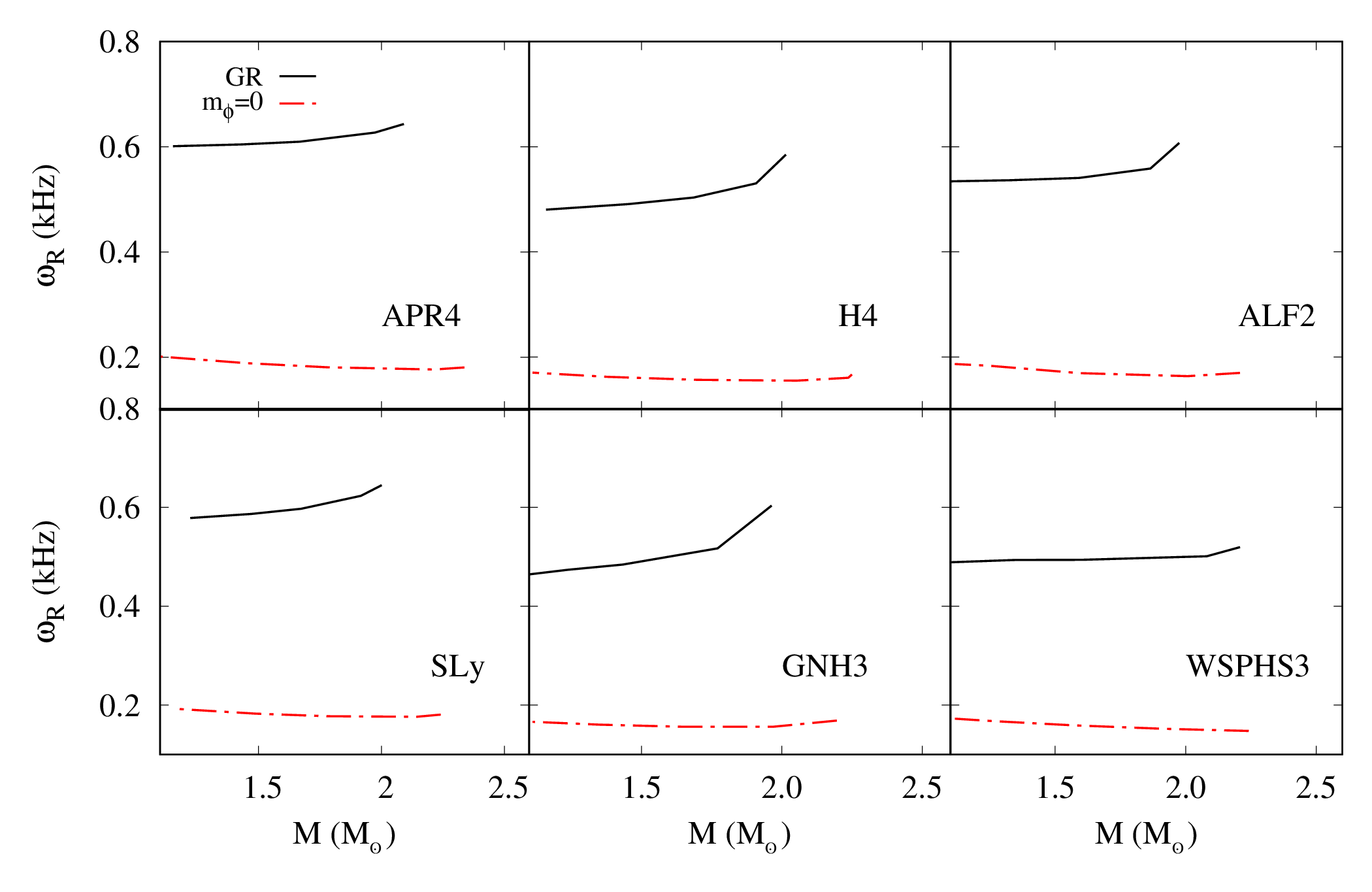}
	\includegraphics[width=.32\textwidth, angle =-90]{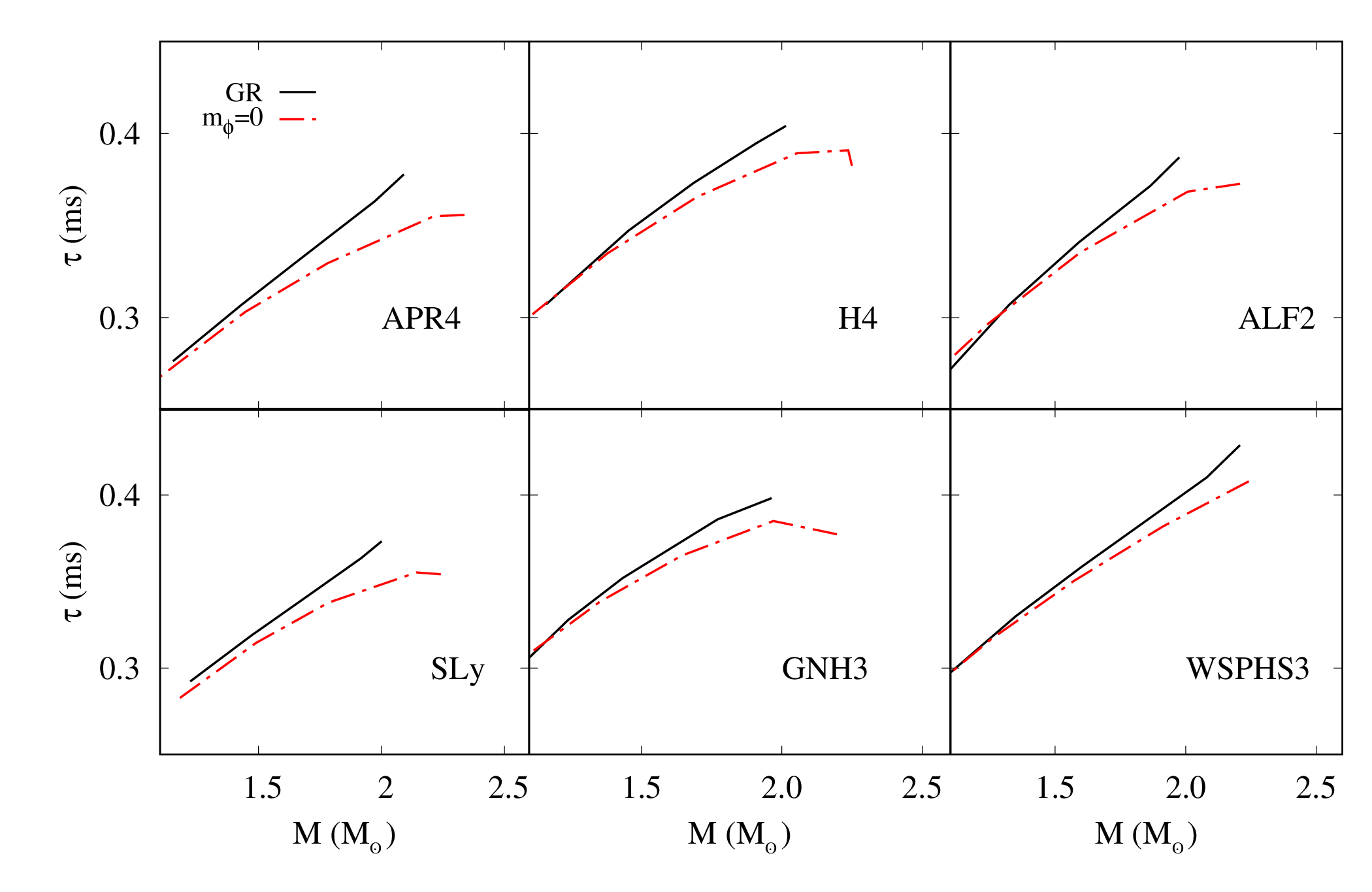}
	\caption{
		Frequency $\omega_R$ in kHz (left) and {damping time $\tau$ in milliseconds (right)} versus neutron star mass $M$ in $M_{\odot}$ for the quadrupole $\phi$-mode.
		{The six panels represent six equations of state, and the color red indicates the massless case with the general relativistic case in black.}
	}
	\label{fig:MR_MI_panel_l2}
\end{figure}

Figure \ref{fig:MR_MI_panel_l2} presents the quadrupole $\phi$-modes for the six selected equations of state. On the left the frequency $\omega_R$ is shown in kHz versus the neutron star mass $M$ in $M_\odot$ and on the right the damping time is shown in milliseconds versus $M$ in $M_\odot$.
In each subfigure, the left panels show the modes for the nuclear matter equations of state, APR4 and SLy, the middel panels for the nucleon-hyperon fluids, H4 and GNH3, and the right panels for the hybrid nuclear-quark matter, ALF2 and WSPHS3.
The red curves exhibit the modes for the massless Brans-Dicke type scalar-tensor theory, while the black curves represent the modes for general relativity with a minimally coupled massless scalar field.

In this massless scalar-tensor theory the frequency is always around 200 Hz or slightly below, whereas in the general relativistic case the frequency is typically much larger, often by up to a factor of three.
Also the frequency increases towards the maximum mass in general relativity, while it mostly decreases in the scalar-tensor case.
The damping time is very close in both theories for small neutron star masses, with deviations getting larger towards the maximum neutron star mass. %
Whereas the damping time rises with increasing mass for both theories, this increase becomes slower in the scalar-tensor case.
Overall, the damping time is mostly in the range of 0.3 to 0.4 ms for both theories.
Note that these are not the only $\phi$-modes in the spectrum of quadrupolar perturbations: many other modes can be found, but these are the ones for which the damping times are larger. 

\subsection{Universal relations for the quadrupole $\phi$-modes}

We now turn to the universal relations for these quadrupole $\phi$-modes. To this end we consider dimensionless quantities (in geometric units) that are formed from the frequency $\omega_R$ and the damping time $\tau$.
In the simplest case these dimensionless quantities are formed with the mass of the star $M$ or the radius of the star $R$.
But they can also involve the radius of gyration $\hat{R}=\sqrt{I/M}$, or the reference frequencies $\omega_o= \sqrt{\frac{3Mc^2}{4{R}^3}}=\frac{c}{M}\sqrt{\frac{3}{4}C^3}$ with the compactness $C=M/R$, or $\hat \omega_o= \sqrt{\frac{3Mc^2}{4\hat{R}^3}}=\frac{c}{M}\sqrt{\frac{3}{4}\eta^3}$ with the generalized compactness $\eta= M/\hat{R} = \sqrt{M^3/I}$, etc.
We then consider these quantities as functions of the compactness $C$, the generalized compactness $\eta$, etc.
When the dimensionless modes lie to good approximation on a single curve for the full set of equations of state, a universal relation is obtained representing the best fit.
We have studied a large set of combinations of dimensionless quantities and exhibit in the following a set of interesting examples for the universal relations for the quadrupole $\phi$-modes. {Here for the quadrupole $\phi$-modes, and also later for the dipole and radial $\phi$-modes, we use a fourth-order polynomial as the fit function.}

\begin{figure}[h!]
	\centering
	\resizebox{0.49\textwidth}{!}{\includegraphics{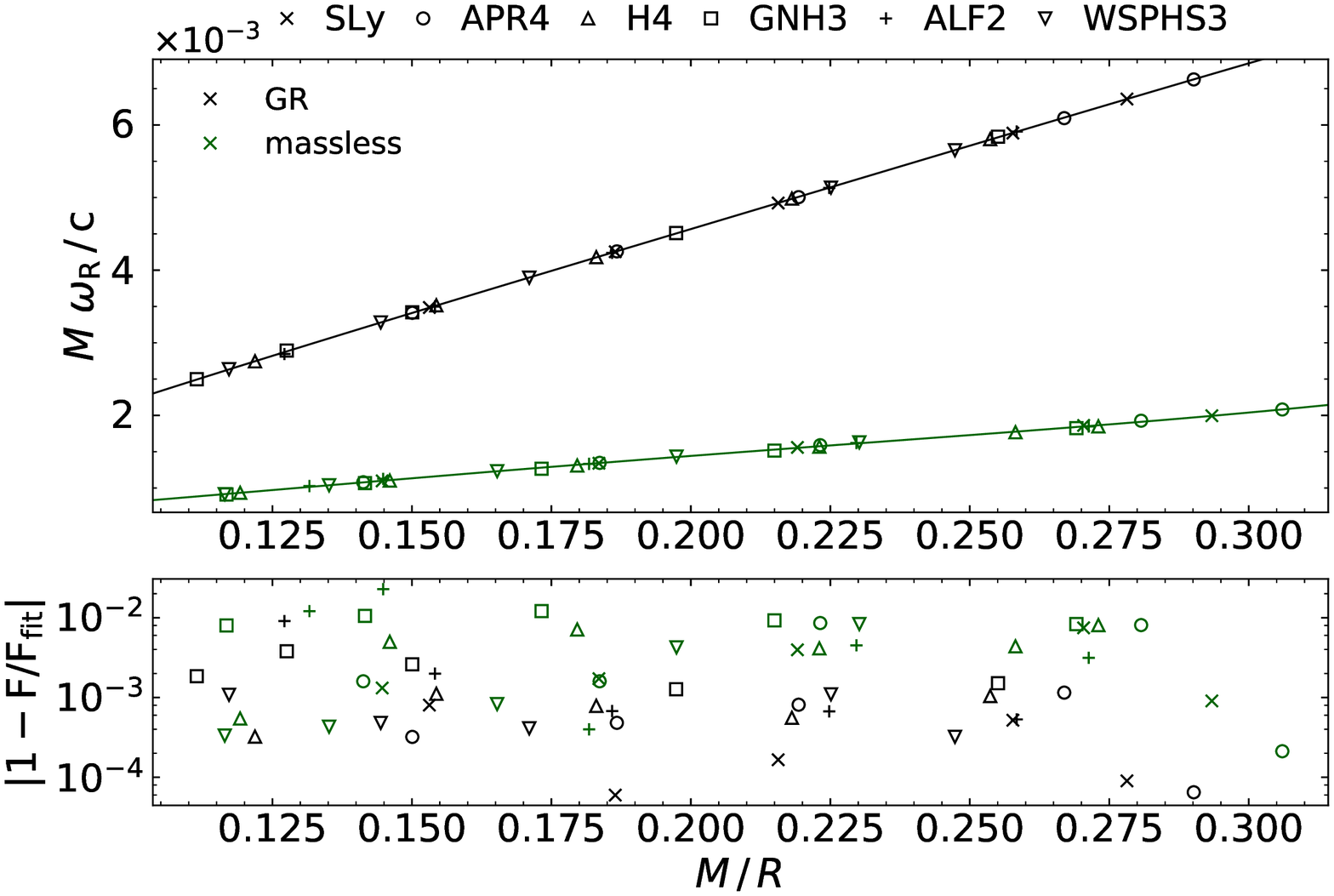}}
	\resizebox{0.49\textwidth}{!}{\includegraphics{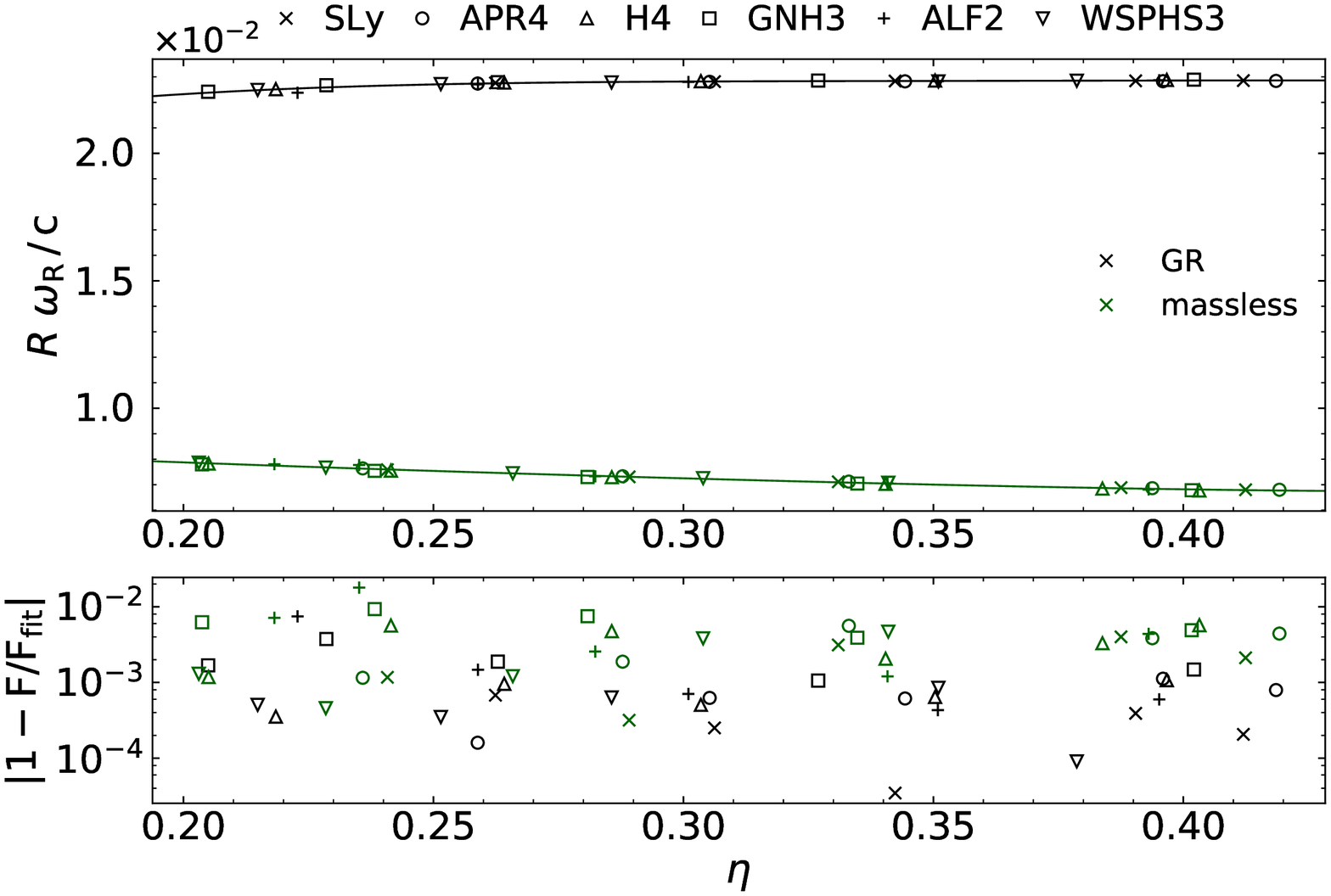}}
	\caption{Quadrupole $\phi$-mode universal relations: dimensionless frequency $M\omega_R/c$ versus compactness $C=M/R$ (left upper panel) and fit errors (left lower panel); dimensionless frequency $R\omega_R/c$ versus generalized compactness $\eta$ (right upper panel) and fit errors (right lower panel).
		The symbols indicate the respective equation of state, the massless scalar-tensor case is shown in green and the general relativistic case in black.}
	\label{fig:my_label2}
\end{figure}

We exhibit in Fig.~\ref{fig:my_label2} a simple set of universal relations for the frequency $\omega_R$ of the modes.
The upper panel of the left figure exhibits the dimensionless frequency $M\omega_R/c$ versus the compactness $C=M/R$ of the star. The symbols identify the respective equation of state, while the colors green and black show the results for the massless scalar-tensor theory and general relativity, respectively.
For both theories rather linear universal relations for the frequency are obtained, that lie far apart and thus are quite distinct.
The lower panel of the figure shows the deviations from the best fit for all of the modes. As already clear from the upper figure, these universal relations are very good, exhibiting a mean error of 0.1\% for general relativity and 0.5\% for the massless scalar-tensor theory.

The right figure in Fig.~\ref{fig:my_label2} employs the radius $R$ instead of the mass for the scaling of the frequency, thus it shows in the upper panel the dimensionless frequency $R\omega_R/c$, but now versus generalized compactness $\eta$, which yields slightly better universal relations (with mean errors 0.1\% and 0.4\% for general relativity and massless scalar-tensor theory, respectively) than the ordinary compactness $C$ (with respective mean errors 0.1\% and 0.5\%).
Again, the relations are almost linear, well separated, and very good.

\begin{figure}[h!]
	\centering
	\resizebox{0.49\textwidth}{!}{\includegraphics{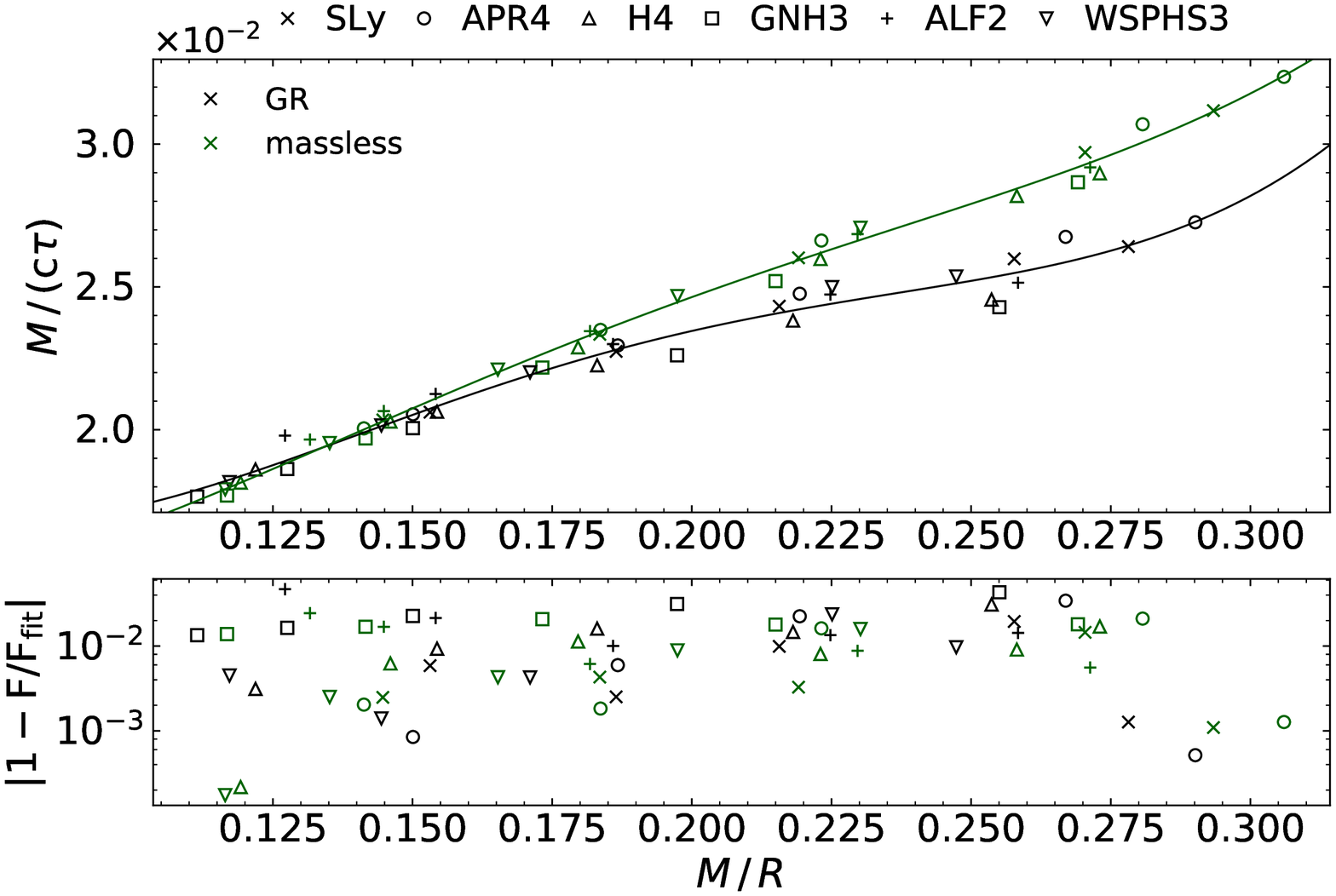}}
	\resizebox{0.49\textwidth}{!}{\includegraphics{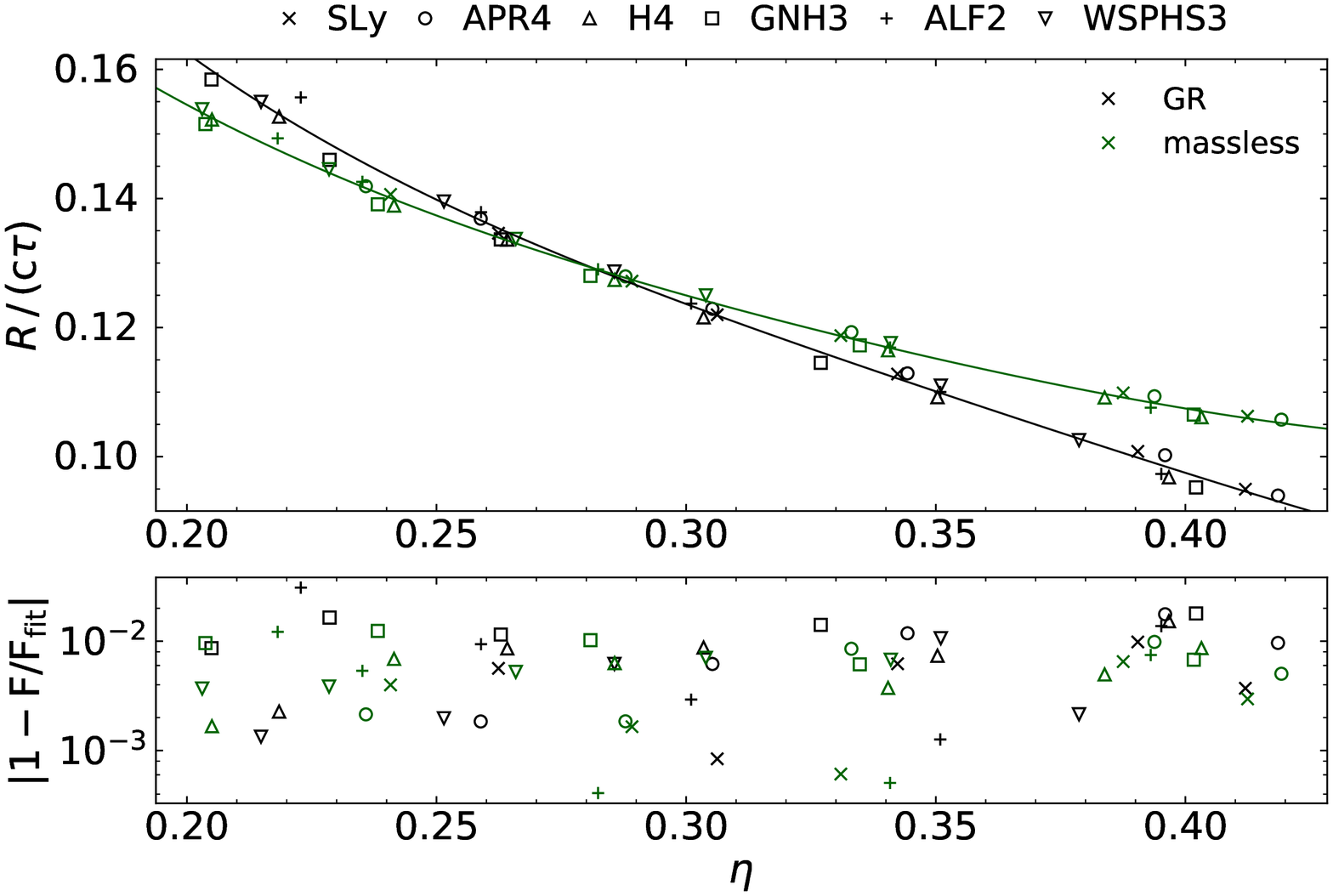}}
	\caption{Quadrupole $\phi$-mode universal relations: dimensionless inverse damping time $M/(c\tau )$ versus compactness $C=M/R$ (left upper panel) and fit errors (left lower panel); dimensionless inverse damping time $R/(c\tau )$ versus generalized compactness $\eta$ (right upper panel) and fit errors (right lower panel).
		The symbols indicate the respective equation of state, the massless scalar-tensor case is shown in green and the general relativistic case in black.}
	\label{fig:my_label3}
\end{figure}

Fig.~\ref{fig:my_label3} shows the corresponding results for the damping time.
Thus the left figure exhibits the universal relations for the dimensionless inverse damping time $M/(c\tau )$ versus the compactness $C=M/R$, while the right figure presents the universal relations for the dimensionless inverse damping time $R/(c\tau )$ versus the generalized compactness $\eta$.
We note that these universal relations for the damping time are not as good as the ones for the frequency, as seen in the lower panels, where again the deviations from the best fits are shown.
Interestingly, they are now better for the massless scalar-tensor case than for general relativity.
However, both theories lead to rather close relations at least in certain ranges of the (effective) compactness, thus making these relations less useful to distinguish between the theories.

\begin{figure}[h!]
	\centering
	\resizebox{0.49\textwidth}{!}{\includegraphics{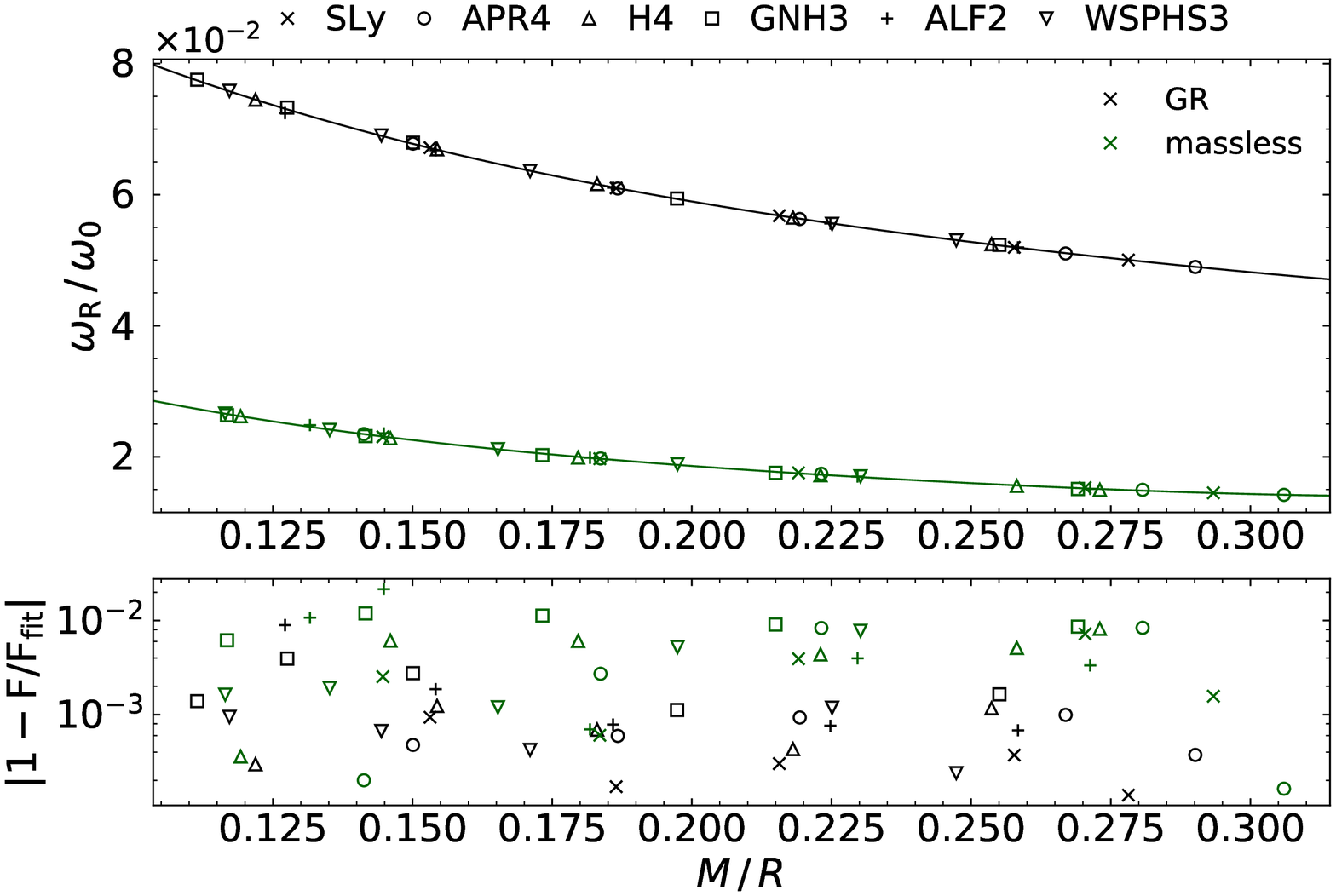}}
	\resizebox{0.49\textwidth}{!}{\includegraphics{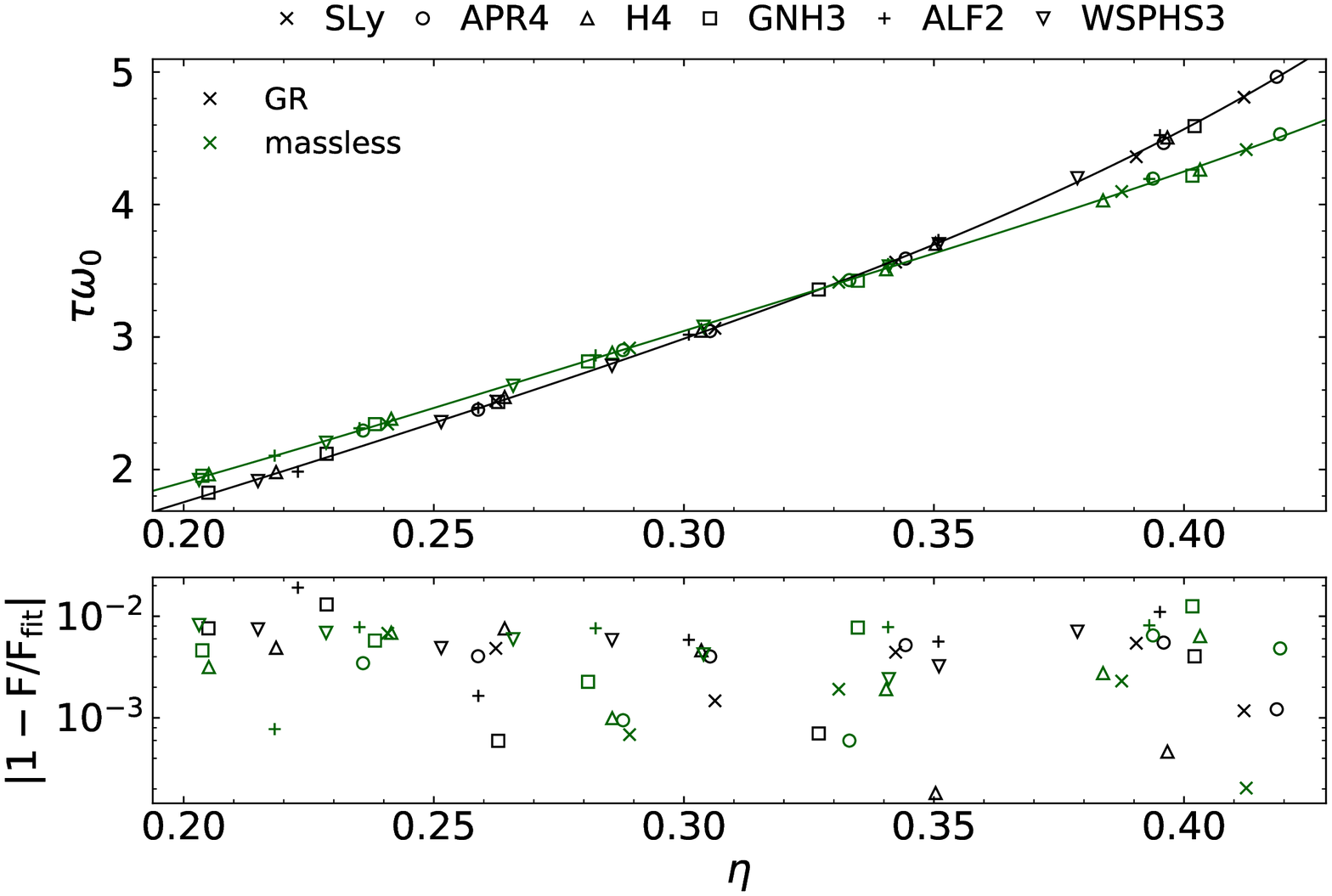}}
	\caption{Quadrupole $\phi$-mode universal relations: dimensionless frequency $\omega_R/\omega_o$ versus compactness $C=M/R$ (left upper panel) and fit errors (left lower panel); dimensionless damping  $\tau\omega_o$ versus generalized compactness $\eta$ (right upper panel) and fit errors (right lower panel).
		The symbols indicate the respective equation of state, the massless scalar-tensor case is shown in green and the general relativistic case in black.}
	\label{fig:my_label4}
\end{figure}

Among the numerous combinations of scaled frequencies and damping times tested, with mean errors displayed in the tables in the appendix, 
we here show another set of very good relations found.
Fig.~\ref{fig:my_label4} exhibits on the left the universal relations for the dimensionless frequency $\omega_R/\omega_o$ versus the compactness $C=M/R$ with mean errors of 0.1\% and 0.6\% for general relativity and the massless scalar-tensor theory, respectively.
Here both relations display a monotonic decrease with increasing $C$, and they differ by a factor of 2 to 3, thus leading to a good discernability of the theories. 
For the dimensionless scaled damping time $\tau\omega_o$ shown in the right figure versus the generalized compactness $\eta$, on the other hand, both relations are mostly very close for general relativity and the massless scalar-tensor theory. Thus although very good (with mean errors of 0.5\% for both) they are not useful to distinguish between the theories.
{Up until this point, the use of solely the dimensionless frequency in the relations is able to discriminate between theories better than solely the dimensionless damping time in the relations.}

\begin{figure}[h!]
	\centering
	\resizebox{0.49\textwidth}{!}{\includegraphics{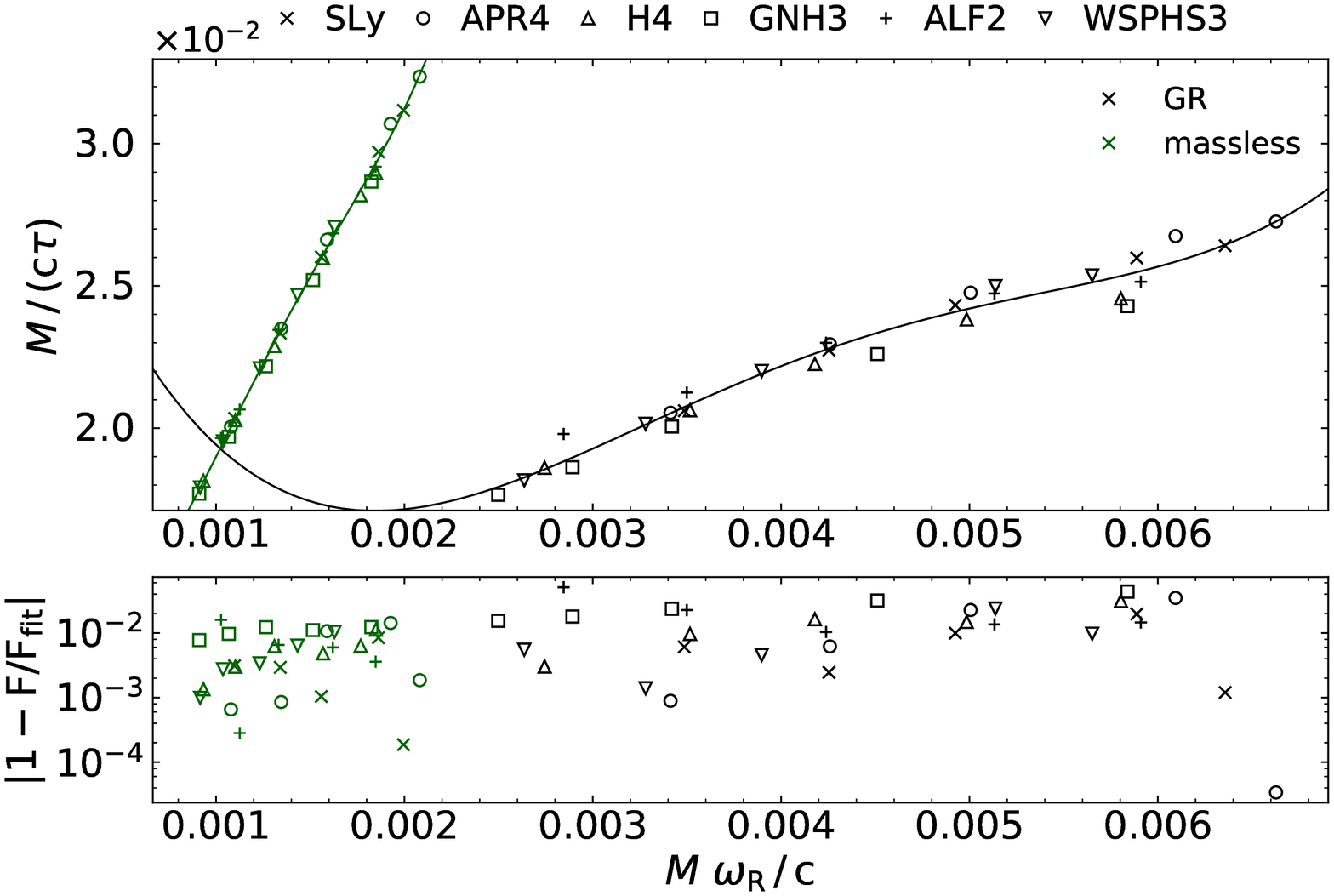}}
	\resizebox{0.49\textwidth}{!}{\includegraphics{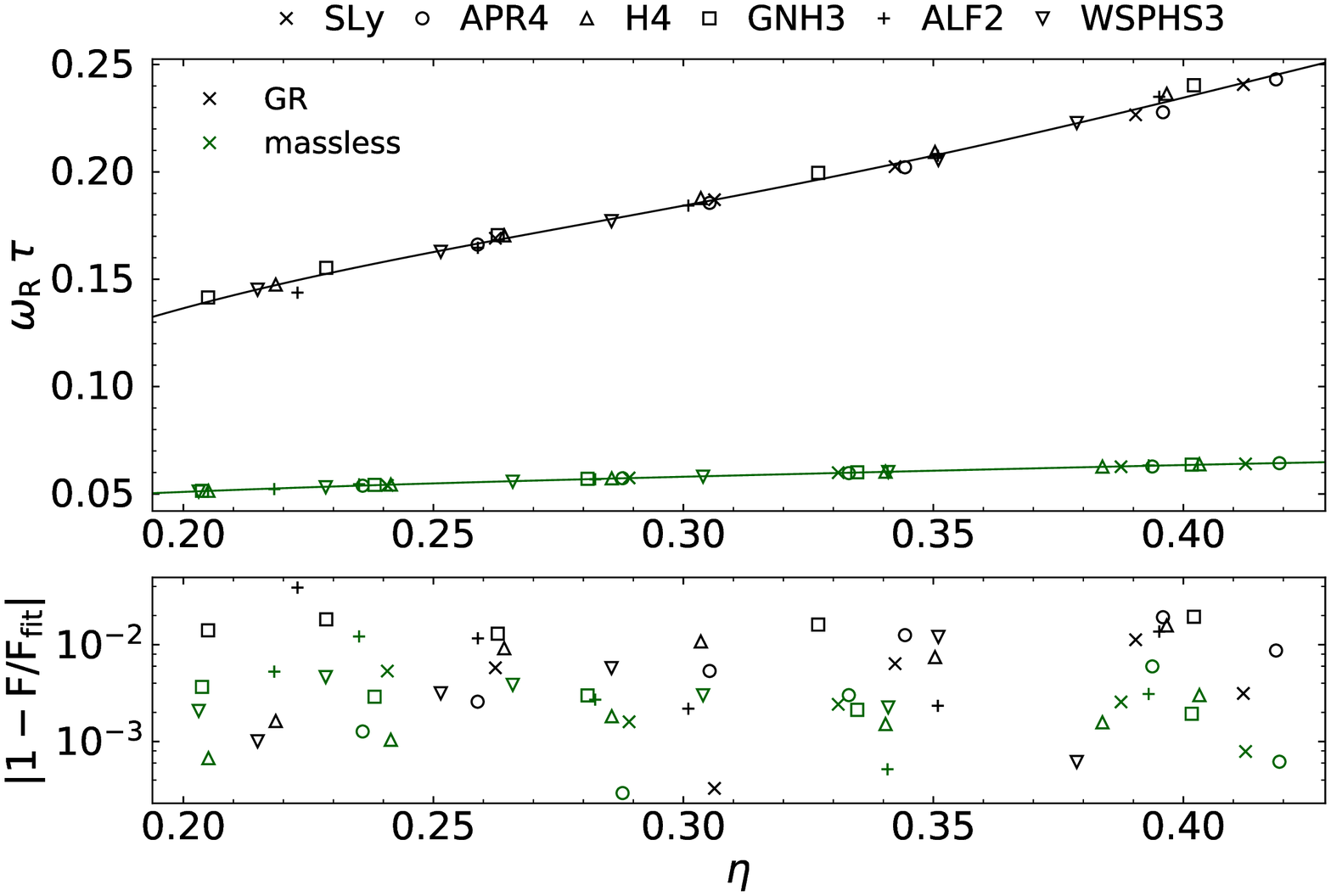}}
	\caption{Quadrupole $\phi$-mode universal relations: dimensionless inverse damping time $M/(c\tau )$ versus dimensionless frequency $M\omega_R/c$ (left upper panel) and fit errors (left lower panel); dimensionless product $\omega_R \tau $ of frequency and damping time versus generalized compactness $\eta$ (right upper panel) and fit errors (right lower panel).
		The symbols indicate the respective equation of state, the massless scalar-tensor case is shown in green and the general relativistic case in black.}
	\label{fig:my_label5}
\end{figure}

The last set of universal relations selected here concerns relations involving both the frequency and the damping time.
We thus show in Fig.~\ref{fig:my_label5} on the left the dimensionless inverse damping time $M/(c\tau )$ versus the dimensionless frequency $M\omega_R/c$ and on the right the dimensionless product $\omega_R \tau $ of the frequency and the damping time versus the generalized compactness $\eta$. 
As seen in the figures the universal relations for general relativity and massless scalar-tensor theory differ considerably, as desired, while their mean errors range from very good for the  massless scalar-tensor theory (0.6\% mean error left figure and 0.3\% right figure) to average for general relativity (1.6\%  left and 1.0\% right).

\section{Dipole $\phi$-modes}\label{App_phi1}

\subsection{Spectrum}

\begin{figure}[h!]
	\centering
	\includegraphics[width=.32\textwidth, angle =-90]{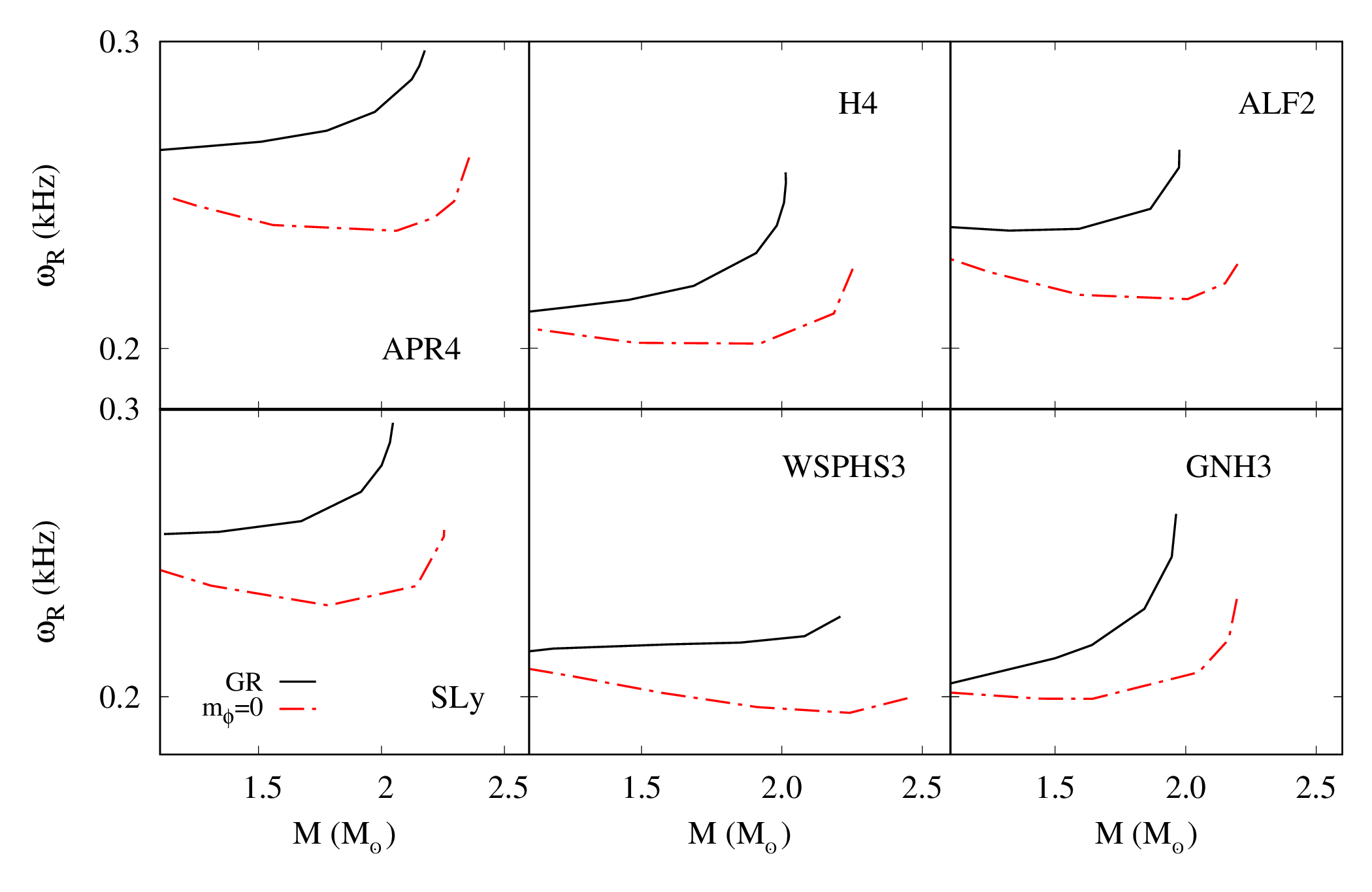}
	\includegraphics[width=.32\textwidth, angle =-90]{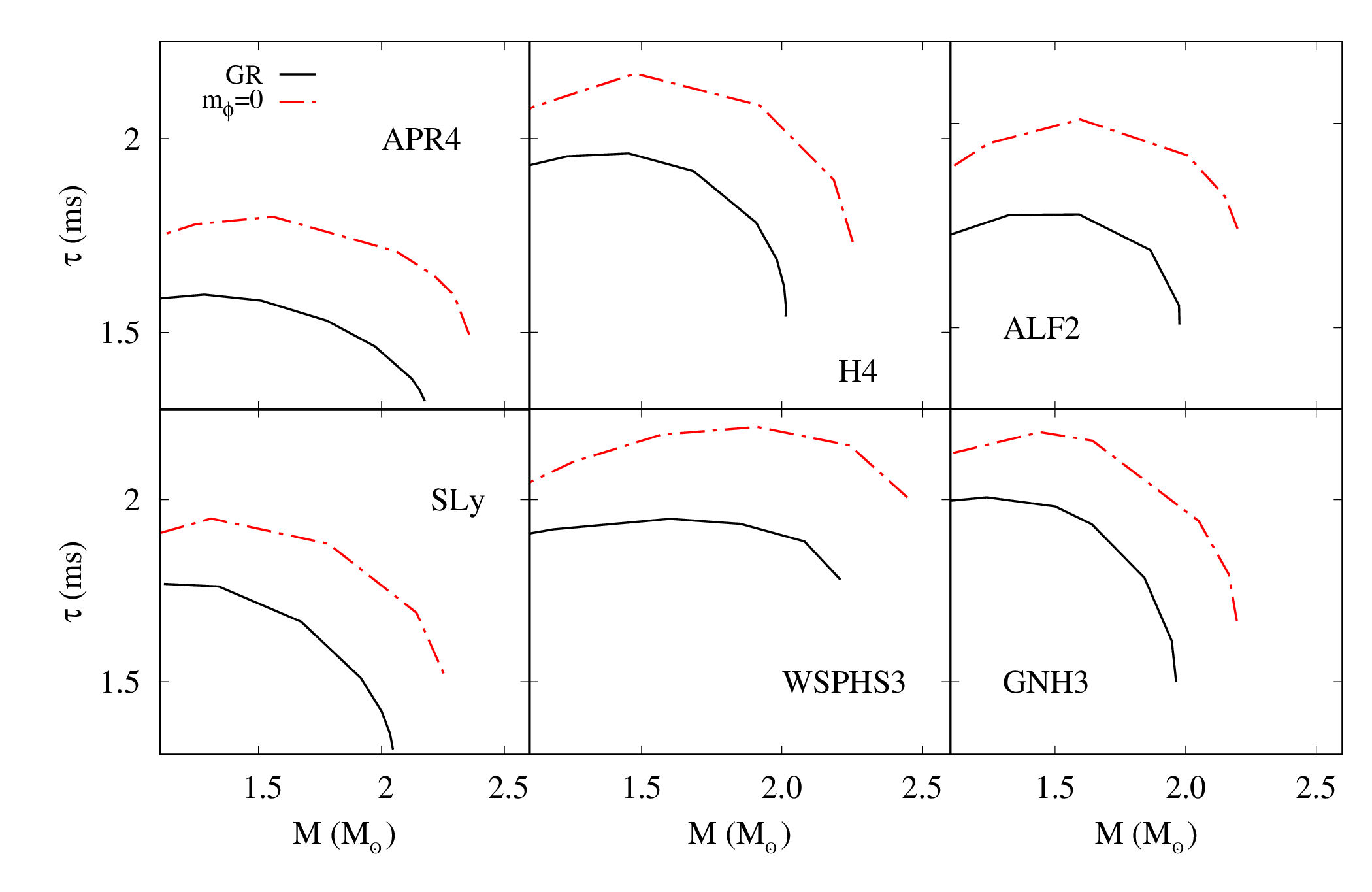}
	\caption{
		Frequency $\omega_R$ in kHz (left) and {damping time $\tau$ in milliseconds (right)} versus neutron star mass $M$ in $M_{\odot}$ for the dipole $\phi$-mode.
		{The six panels represent six equations of state, and the color red indicates the massless case with the general relativistic case in black.}
	}
	\label{fig:MR_MI_panel_l1}
\end{figure}

We now turn to the dipole $\phi$-modes of the neutron stars.
We exhibit these $\phi$-modes for the chosen set of equations of state in Fig.~\ref{fig:MR_MI_panel_l1}, with frequency $\omega_R$ in kHz versus the neutron star mass in $M_\odot$ on the left and the damping time in milliseconds on the right.
The black curves present the results for general relativity with a minimally coupled massless scalar field, and the red curves show the results for the massless scalar-tensor theory. 
The frequency of the dipole $\phi$-modes is always below 300 Hz, which is quite a bit lower than for the dipole F-modes obtained before \citep{Blazquez-Salcedo:2022pwc}.
For the dipole $\phi$-modes general relativity leads to larger frequencies than the massless scalar-tensor theory. 
In general the frequency tends to increase for configurations close to the maximum mass.
The damping time $\tau$ is typically less than 2 milliseconds for the general relativistic case, and the introduction of the massless scalar-tensor theory has the overall effect of increasing the damping time of the $\phi$-mode. 
In general the damping time tends to decrease for configurations close to the maximum mass.

\subsection{Universal relations for the dipole $\phi$-modes}\label{App_phi2}

We now address the universal relations for the dipole $\phi$-modes for the two considered theories, the massless scalar-tensor theory and general relativity with a minimally coupled scalar field.
We proceed as for the quadrupole $\phi$-mode discussed in the previous section.

\begin{figure}[h!]
	\centering
	\includegraphics[width=.49\textwidth, angle =0]{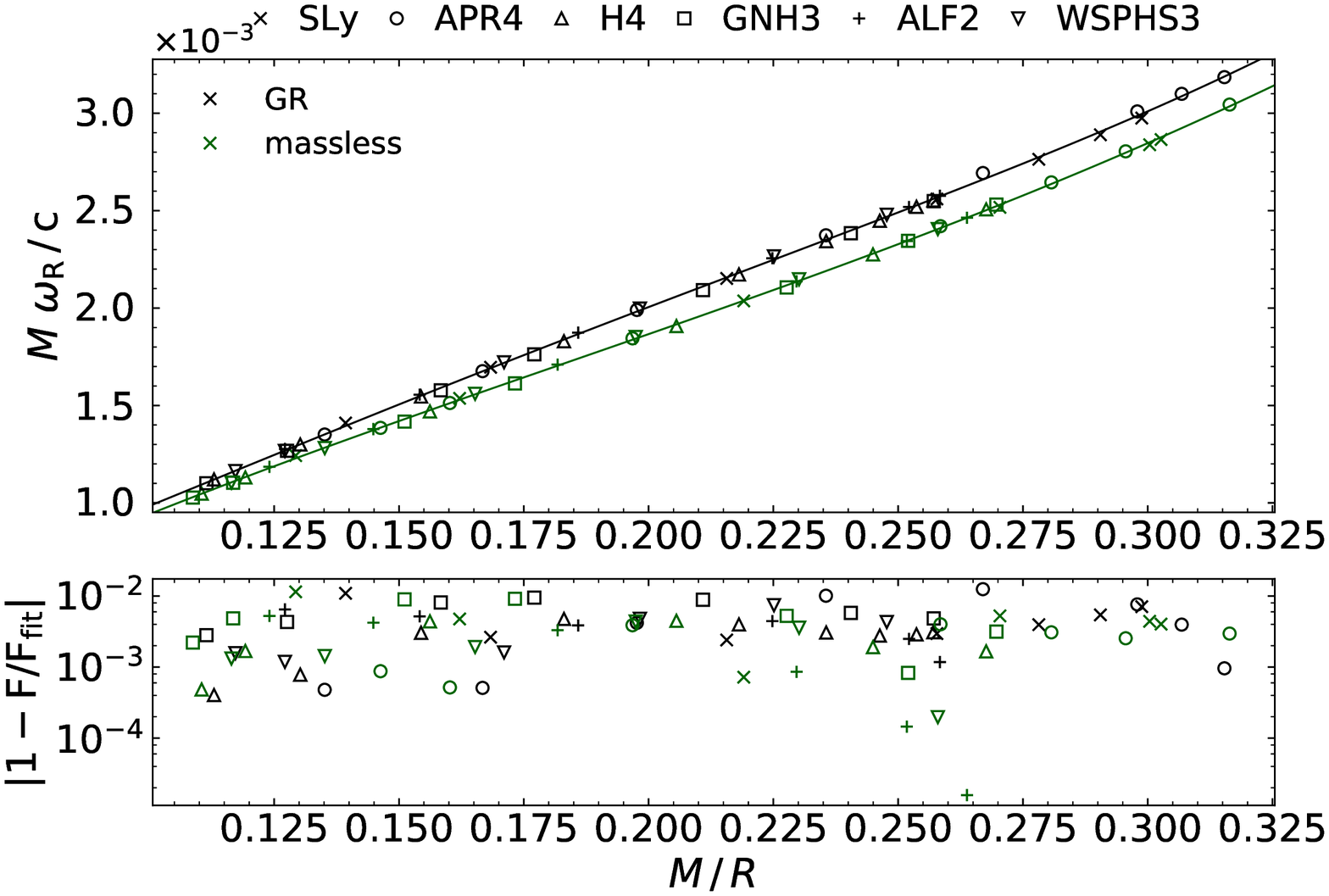} 
	\includegraphics[width=.49\textwidth, angle =0]{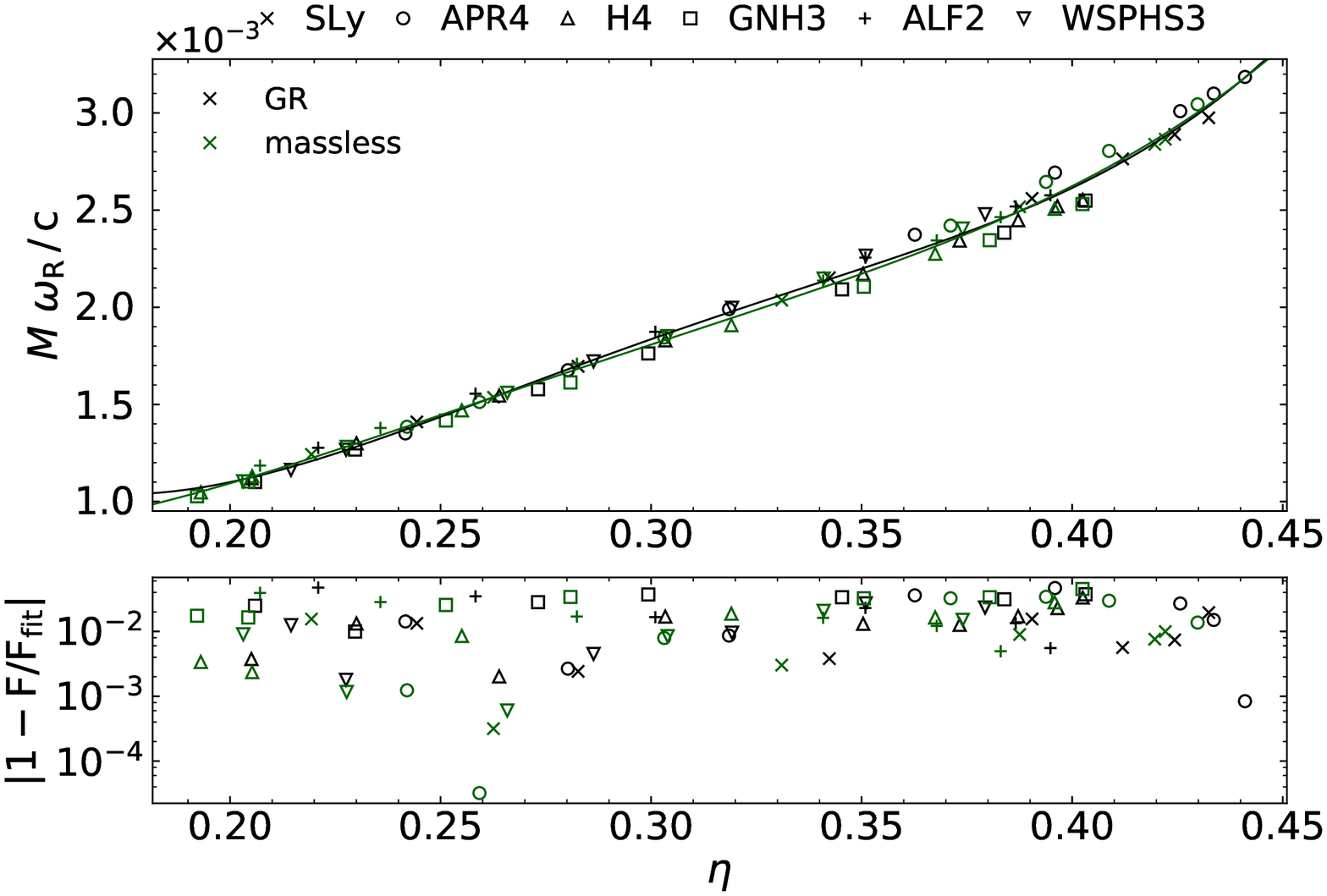}
	\caption{{Dipole $\phi$-mode universal relations: dimensionless frequency $M\omega_R/c$ (upper panels) and fit errors (lower panels) versus compactness $C=M/R$ (left panels); versus generalized compactness $\eta$ (right panels).
			The symbols indicate the respective equation of state, and the color green the massless case with the general relativistic case in black.}
	}
	\label{fig:uni_rel_MOmegaR_scalar_l1_error}
\end{figure}

\begin{figure}[h!]
	\centering
	\includegraphics[width=.49\textwidth, angle =0]{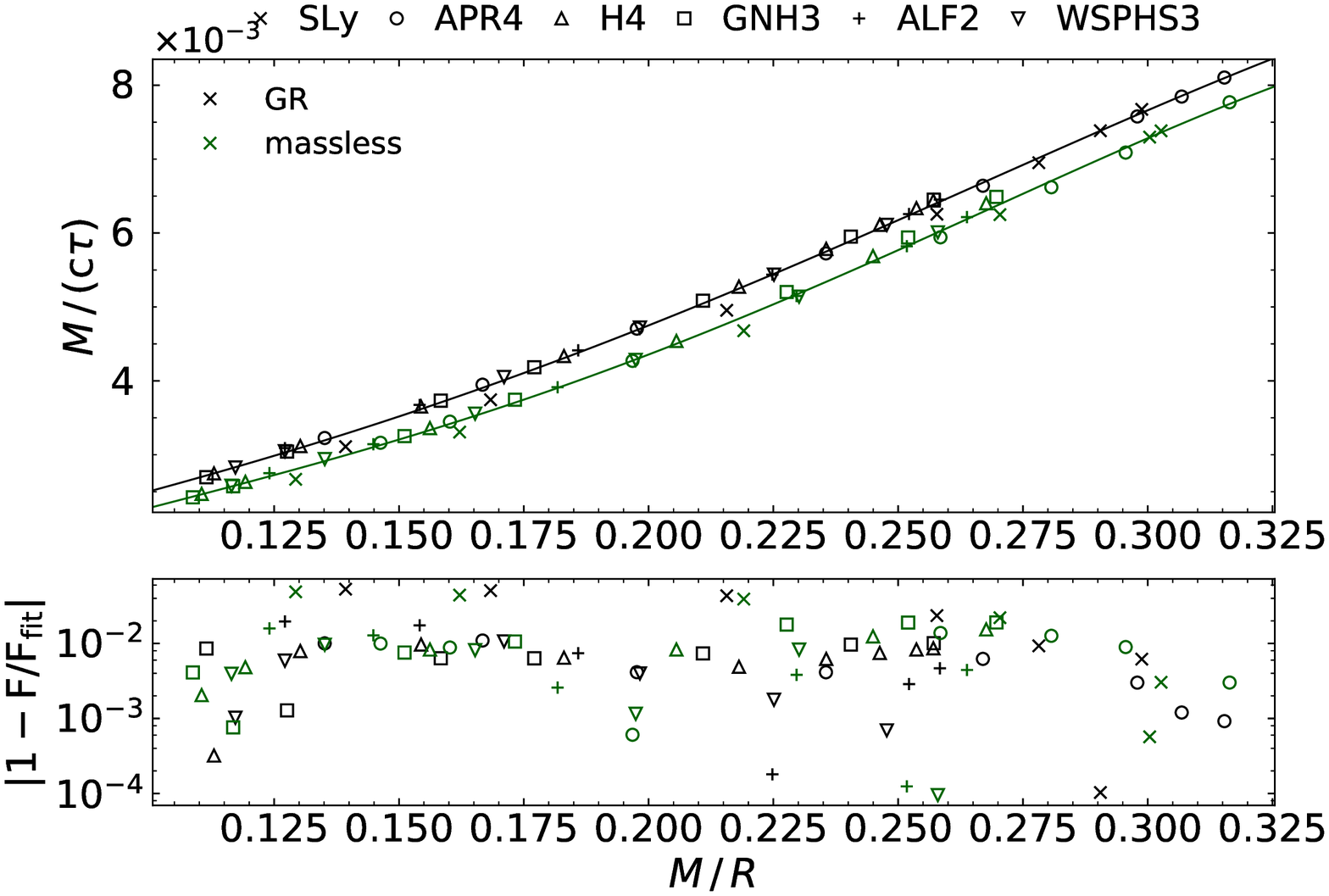}
	\includegraphics[width=.49\textwidth, angle =0]{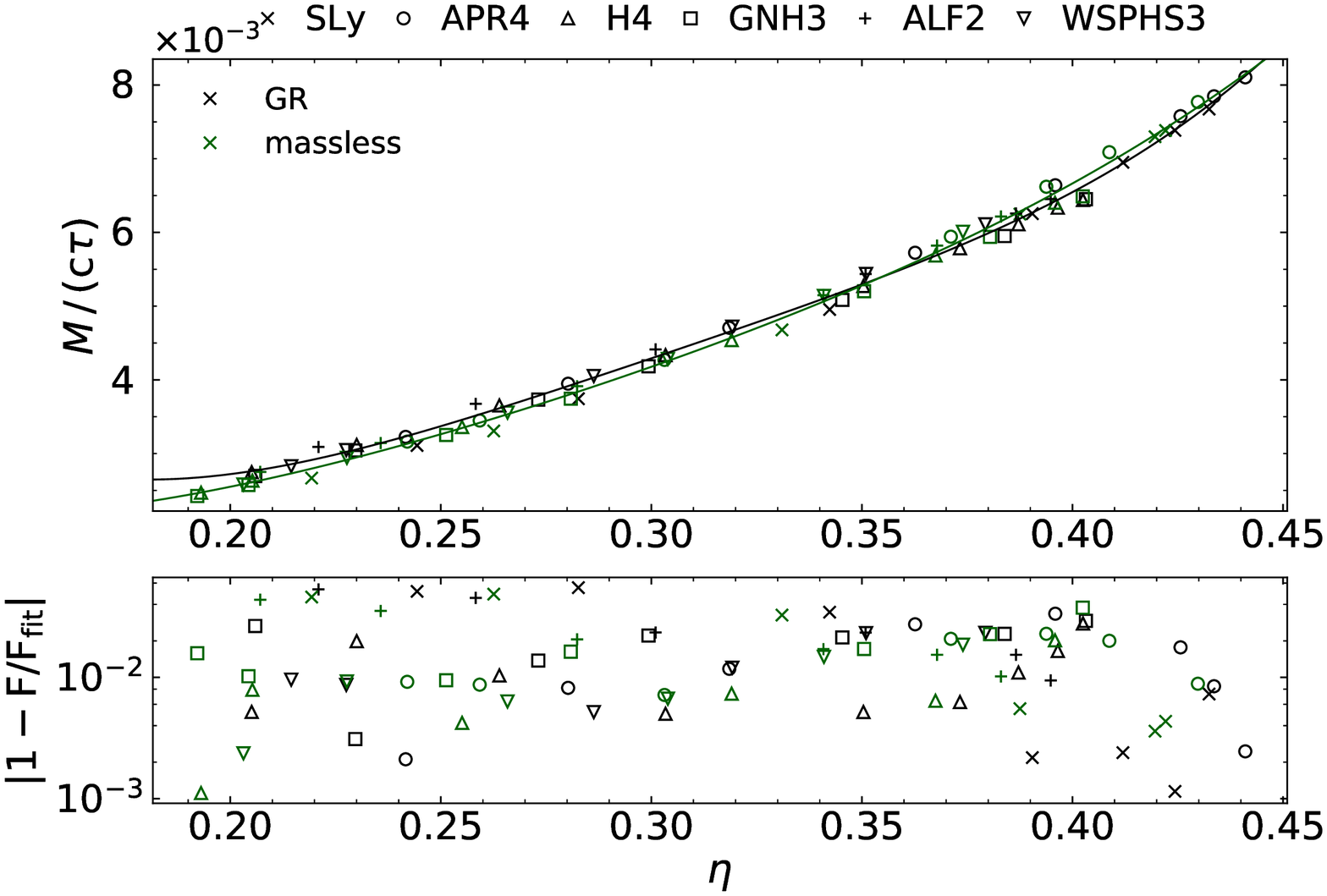}
	\caption{{Dipole $\phi$-mode universal relations: dimensionless inverse damping time $M/(c\tau)$ (upper panels) and fit errors (lower panels) versus compactness $C=M/R$ (left panels); versus generalized compactness $\eta$ (right panels).
			The symbols indicate the respective equation of state, and the color green the massless case with the general relativistic case in black.}
	}
	\label{fig:uni_rel_MOmegaI_scalar_l1_error}
\end{figure}

In Figure \ref{fig:uni_rel_MOmegaR_scalar_l1_error} we show the dimensionless frequency $M\omega_R/c$ scaled with the neutron star mass $M$ versus the compactness $C$ on the left and $M\omega_R/c$ versus the generalized compactness $\eta$ on the right.
The color green indicates the massless theory, and the results for general relativity are shown in black.
The lower panels contain as before the associated fit errors.
This simple scaling with the mass works very well for the dimensionless frequency $M \omega_R/c$ $-$ compactness $C$ relations, with low mean errors of {0.4\% and 0.3\% for general relativity and the massless theories respectively}.
{These relations are far better than the relations involving the generalized compactness with the respective mean errors of 1.8\% and 1.6\%.}
Moreover, in the case of the generalized compactness the curves for the massless theory and general relativity are very close.

Figure \ref{fig:uni_rel_MOmegaI_scalar_l1_error} represents the corresponding figure for the damping time $\tau$, i.e., the dimensionless inverse damping time $M/c \tau $ is shown versus the compactness $C$ (left) and generalized compactness $\eta$ (right).
The fit reveals, that the errors are larger for the damping time than for the frequency, with mean errors of 1.0\% and 1.1\% in the case of compactness and 1.8\% and 1.6\% in the case of generalized compactness.
\begin{figure}[h!]
	\centering
	\includegraphics[width=.49\textwidth, angle =0]{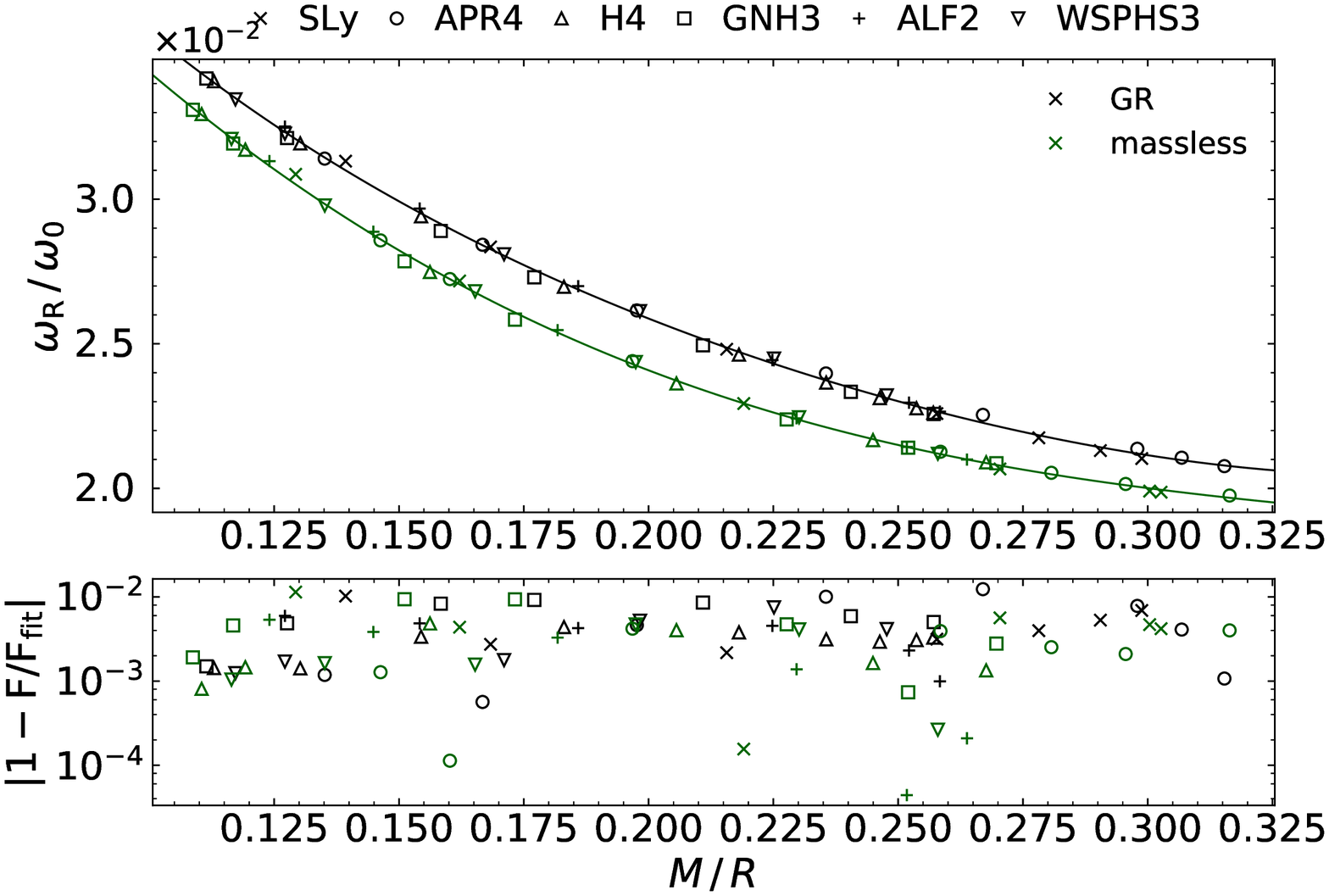}
	\includegraphics[width=.49\textwidth, angle =0]{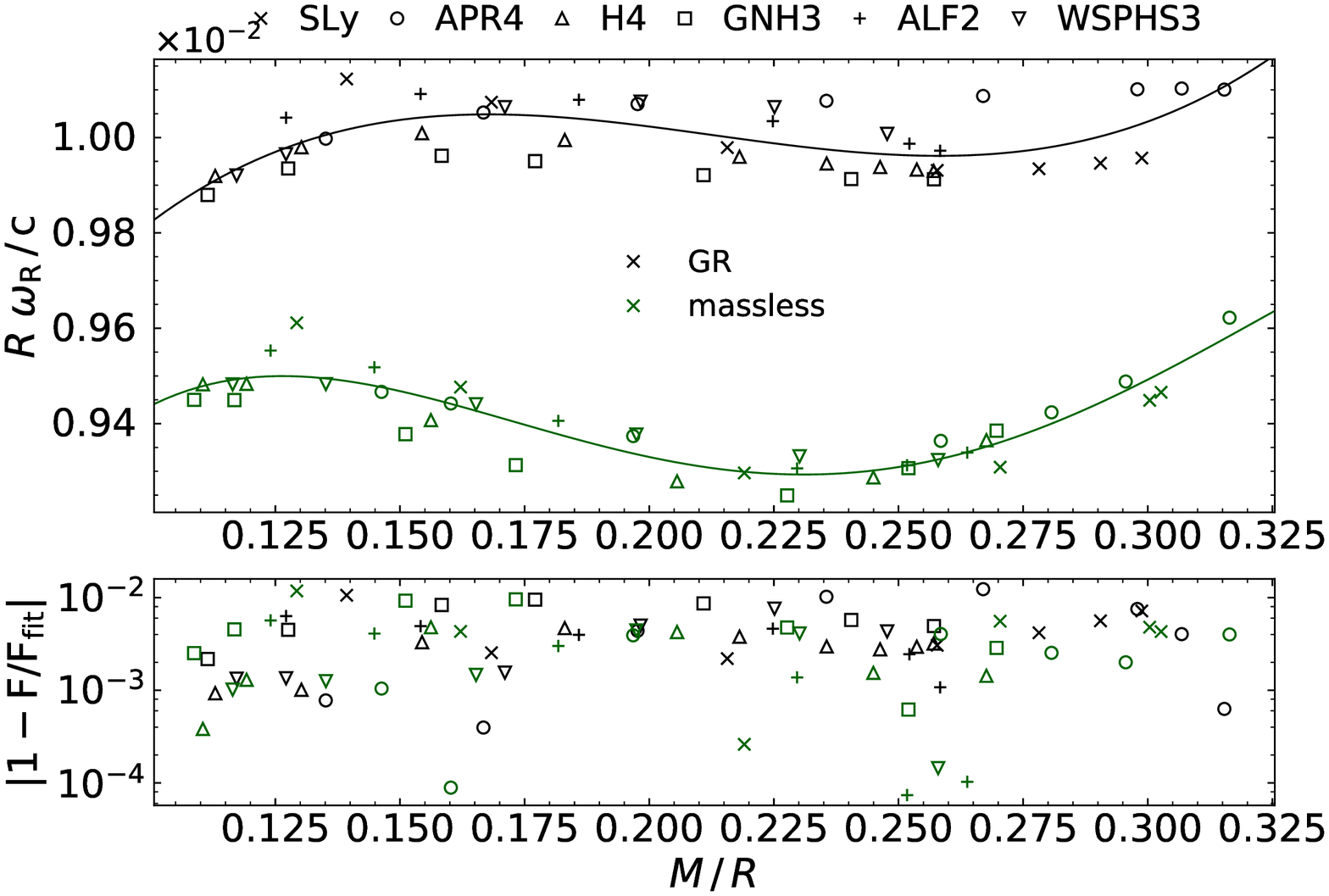}
	\caption{{Dipole $\phi$-mode universal relations: dimensionless frequencies {$\omega_R/\omega_o$} 
			(left upper panel) 
			and $R\omega_R/c$ (right upper panel), 
			and their fit errors (lower panels) versus compactness $C=M/R$.
			The symbols indicate the respective equation of state and the color green the massless case with the general relativistic case in black.}
	}
	\label{fig:uni_rel_c_scalar_l1_error_no_log}
\end{figure}

Fig.~\ref{fig:uni_rel_c_scalar_l1_error_no_log} (left) shows the universal relations for the dimensionless frequency $\omega_R/\omega_o$ versus the compactness $C$. 
{These relations are very good and also distinct, showing means errors of 0.4\% for GR and 0.3\% for the massless theory.} 
Another set of very good universal relations for the frequency $\omega_R$ is shown in Fig.~\ref{fig:uni_rel_c_scalar_l1_error_no_log} (right), where the dimensionless $R\omega_R/c$ is considered as a function of the compactness $C$. {Again the mean errors are very small with 0.4\% and 0.3\% for general relativity and the massless scalar-tensor theory, respectively.}

\begin{figure}[h!]
	\centering
	\includegraphics[width=.49\textwidth, angle =0]{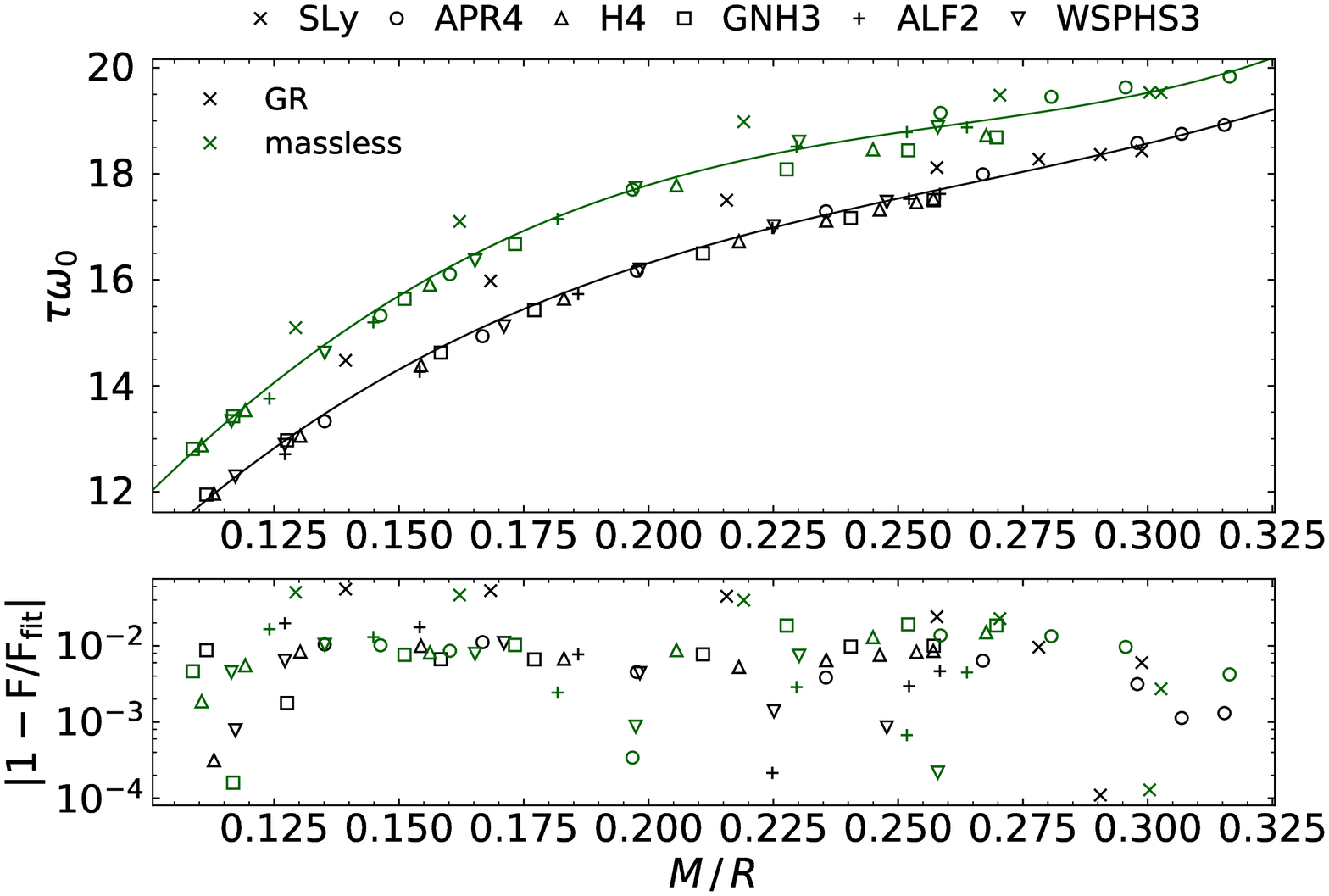}
	\includegraphics[width=.49\textwidth, angle =0]{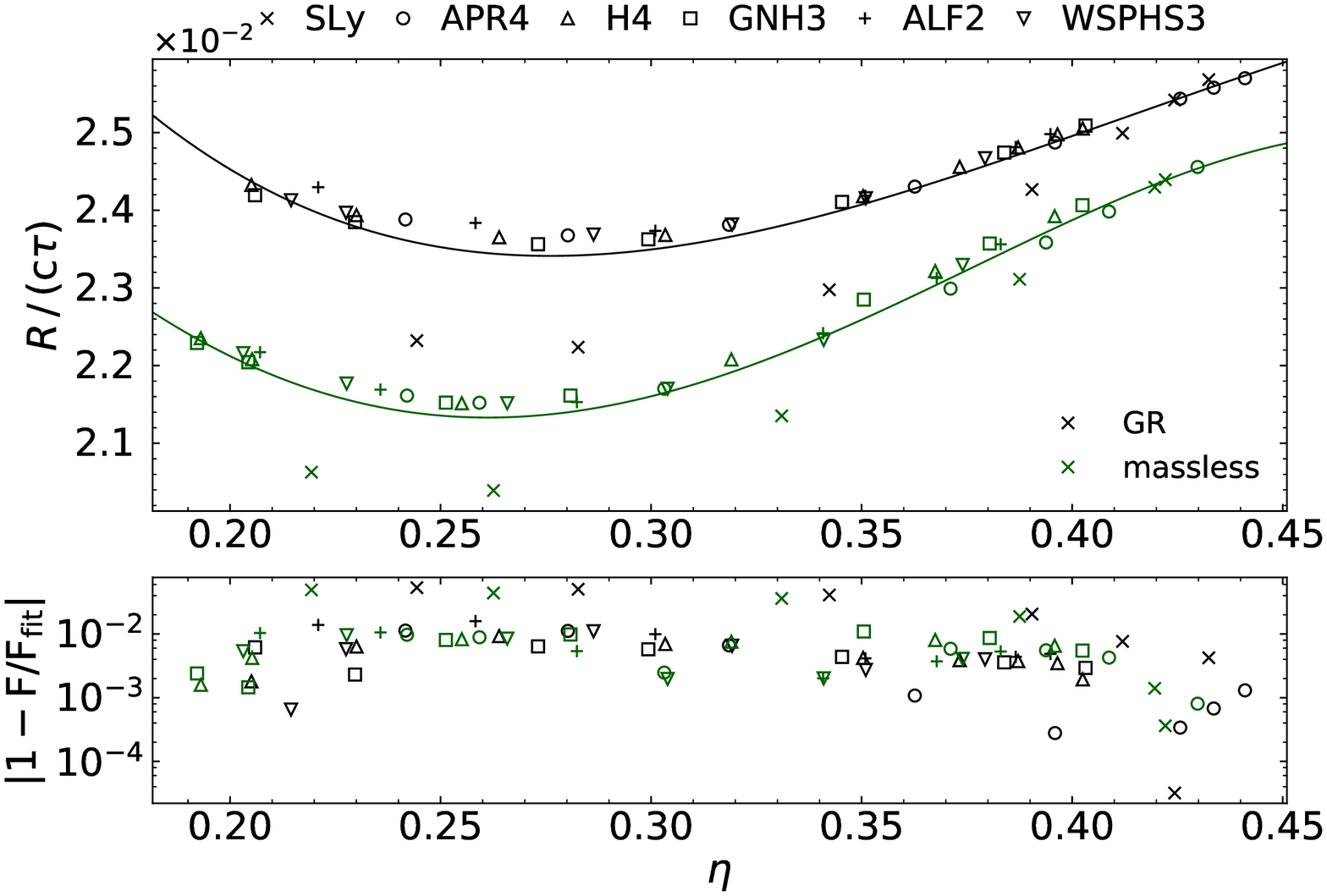}
	\caption{{
			{Dipole $\phi$-mode universal relations: dimensionless damping time $\tau\omega_o$ versus compactness $C=M/R$ (left upper panel) and fit errors (left lower panel); and dimensionless inverse damping time $R/(c\tau)$ versus generalized compactness $\eta$ (right upper panel) and fit errors (right lower panel).
				The symbols indicate the respective equation of state, and the color green the massless case with the general relativistic case in black.}}
	}
	\label{fig:uni_rel_tau_scale_l1_phi}
\end{figure}

When considering the damping time $\tau$, the quality of the universal relations decreases again.
We exhibit in Fig.~\ref{fig:uni_rel_tau_scale_l1_phi} (left) the dimensionless damping time $\tau\omega_o$ versus the compactness $C$. 
Scaling with $\hat{\omega}_0$ results in a worse fit. 
{In Fig.~\ref{fig:uni_rel_tau_scale_l1_phi} (right) the dimensionless quantity $R/(c\tau)$ is shown versus the generalized compactness $\eta$. 
	For this dimensionless quantity the generalized compactness yields a better relation than it is against the compactness. 
	The relations shown for $\tau$ are all equally meaningful, with Fig.~\ref{fig:uni_rel_tau_scale_l1_phi} (right) showing a slightly better fit.} 
Further dimensionless quantities that have been tested are $R/(cC\tau)$, $R/(cC^3\tau)$, $R\omega_R/(cC)$, $R\omega_R/(cC^3)$, and $\omega_R\tau$, versus the compactness and the generalized compactness. Those versus the compactness always lead to an improvement. Relations for the dimensionless damping time $M/(c\tau)$ versus the dimensionless frequency $M\omega_R/c$ have also been tested.  In addition, we have also examined the relations for $\hat{R}\omega_R/c$ and $\hat{R}/c\tau$ with respect to the generalized compactness $\eta$ separately, but none of them provides further improvements in the errors of the relations or in the discernability between the theories. In fact, similar to what is observed in Fig.~\ref{fig:uni_rel_MOmegaI_scalar_l1_error}, when the proposed quantities are considered versus the generalized compactness $\eta$, the splitting of these relations with respect to the theories tends to diminish.

\section{Radial $\phi$-modes}

\subsection{Spectrum}

\begin{figure}[h!]
	\centering
	\includegraphics[width=.32\textwidth, angle =-90]{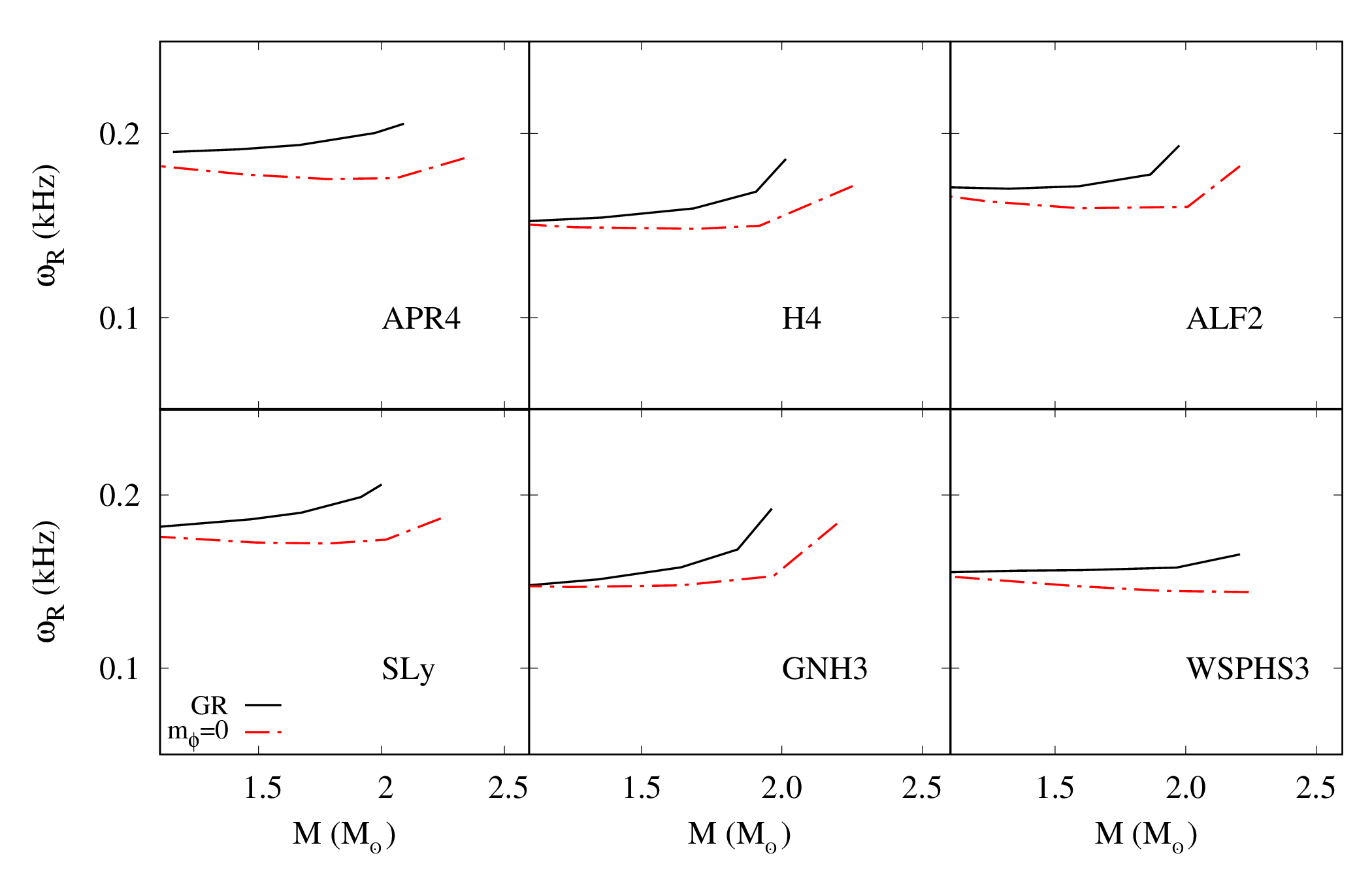}
	\includegraphics[width=.32\textwidth, angle =-90]{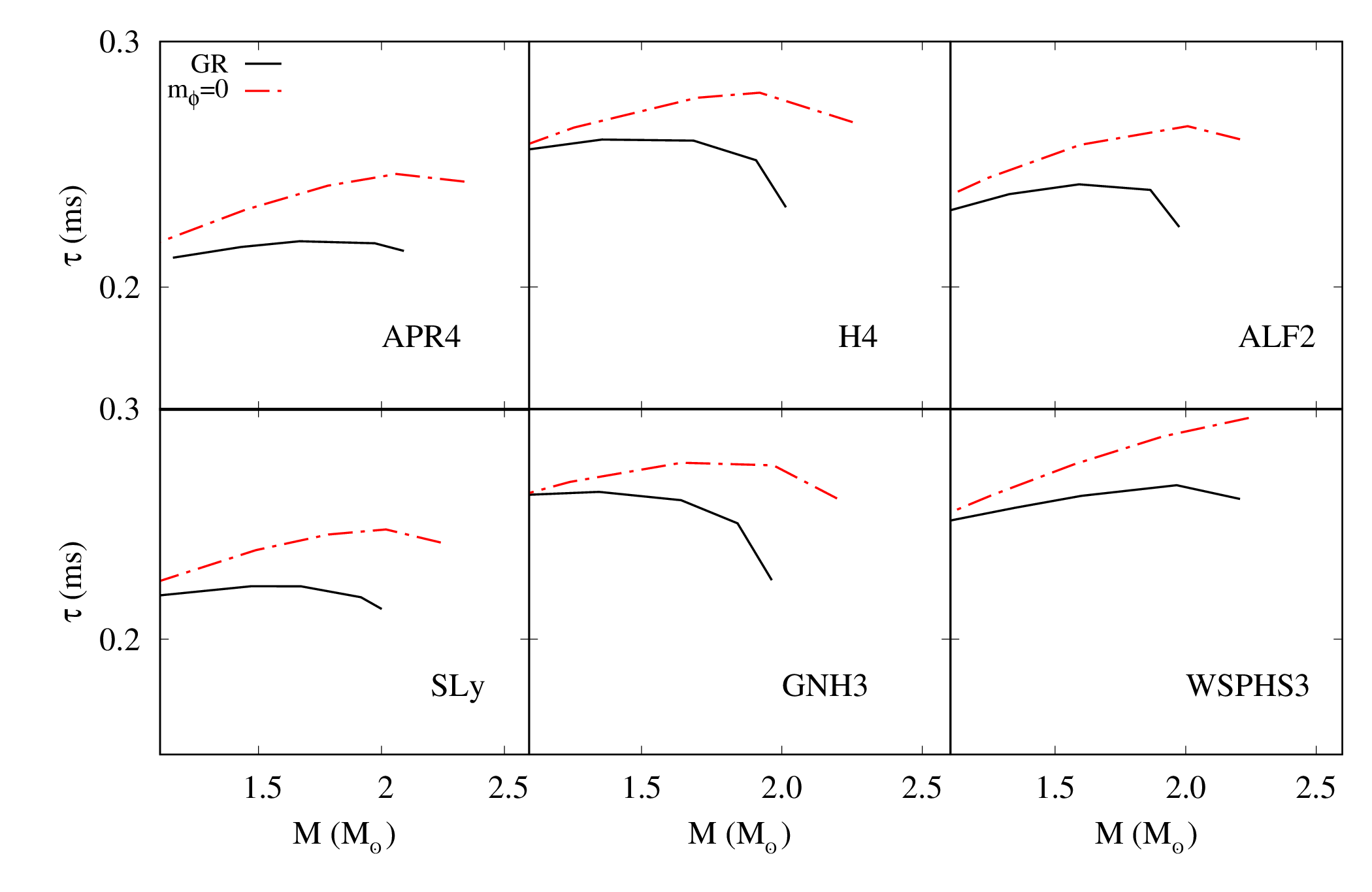}
	\caption{
		Frequency $\omega_R$ in kHz (left) and {damping time $\tau$ in milliseconds (right)} versus neutron star mass $M$ in $M_{\odot}$ for the radial $\phi$-mode.
		{The six panels represent six equations of state, and the color red indicates the massless case with the general relativistic case in black.}
	}
	\label{fig:MR_MI_panel_l0}
\end{figure}

We exhibit the sets of radial $\phi$-modes for general relativity and the massless scalar-tensor theory in Fig.~\ref{fig:MR_MI_panel_l0}. 
The frequency $\omega_R$ of the modes is shown in the left figure, and is located mostly in the range of 100 - 200 Hz.
The damping time $\tau$ of the modes is on the order of 0.2 - 0.3 milliseconds, as seen in the right figure. Although these are not the only scalar-led modes in the spectrum of spherical perturbations, these are the modes with the highest damping time and best numerical precision in the shooting method we employ \footnote{In some cases the numerical calculations produce modes with shorter frequencies (of the order of 50 Hz).}.

For the less massive neutron stars the frequency and the damping time are very similar for both theories, but deviate sizeably towards the maximum mass of the stars, with the general relativistic frequency larger and the damping time smaller than their counterparts in the massless scalar-tensor theory.

\begin{figure}[h!]
	\centering
	\includegraphics[width=.3\textwidth, angle =-90]{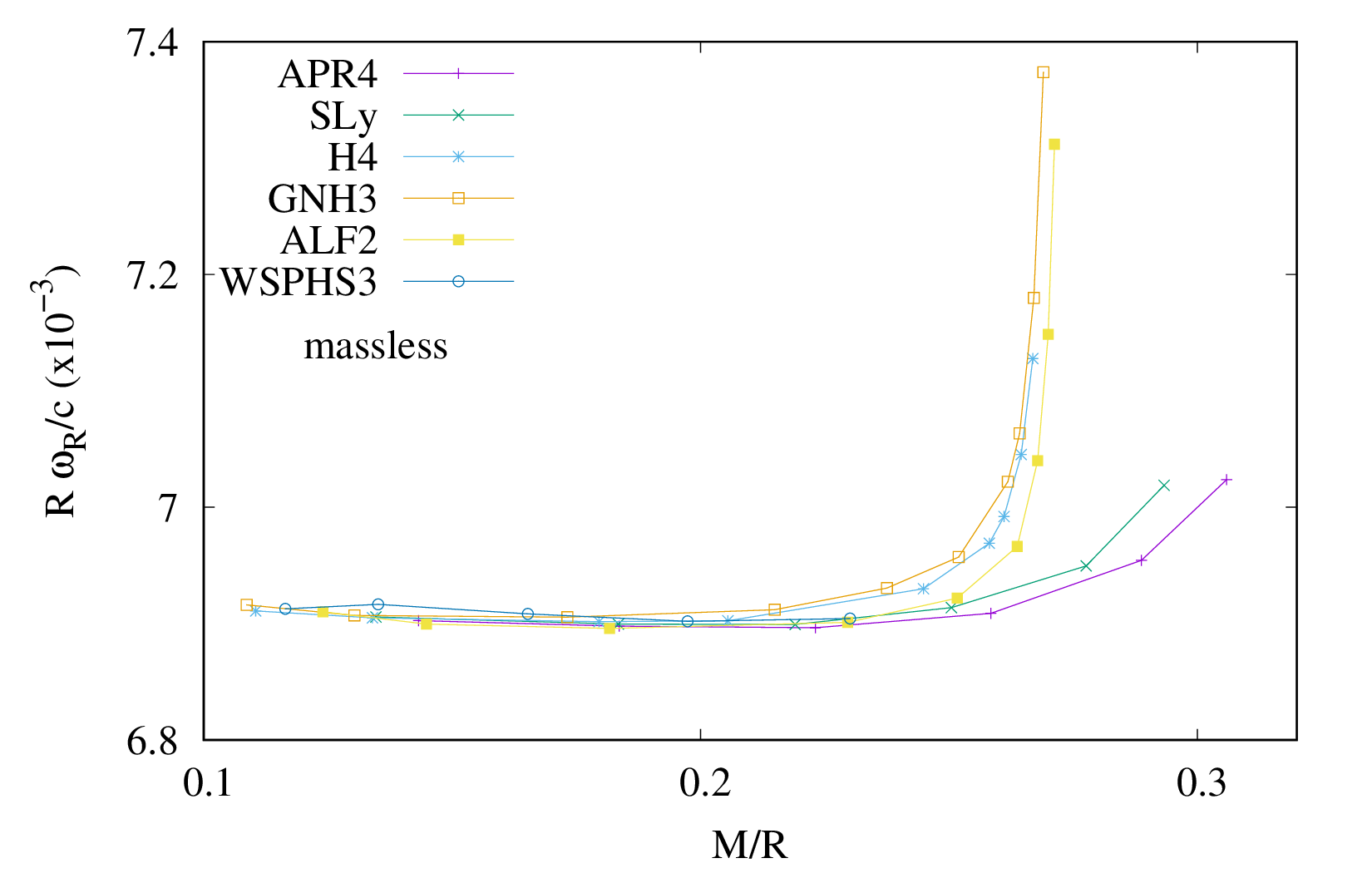}
	\includegraphics[width=.3\textwidth, angle =-90]{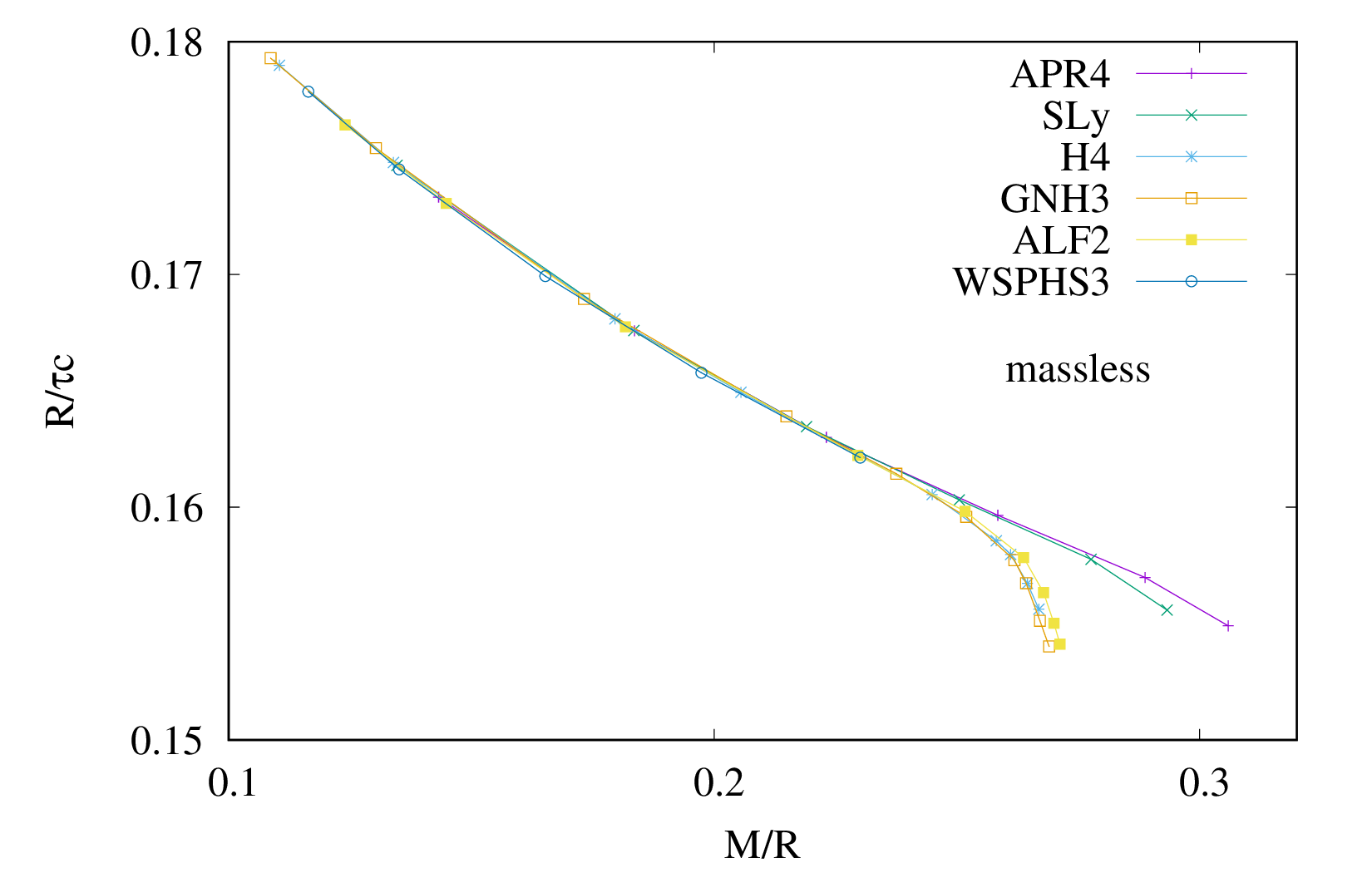}
	\caption{Scaled radial $\phi$-modes for the massless scalar-tensor theory: dimensionless frequency $R\omega_R/c$ versus compactness $C=M/R$ (left); dimensionless inverse damping time $R/(c\tau )$ versus compactness $C$ (right). 
		The colors indicate the different equations of state.}
	\label{fig:my_label12a}
\end{figure}

In Fig.~\ref{fig:my_label12a} we show the scaled frequency $R\omega_R/c$ (left) and the scaled inverse damping time $R/(c\tau )$ versus the compactness $C$ for the massless scalar-tensor theory.
The figure highlights that the scaled quantities are very close to each other for the different equations of state except in a region close the respective maximum mass.
At the maximum mass the instability sets in, found in the $l=0$ sector of the theory, in the fundamental fluid F-mode \citep{Blazquez-Salcedo:2020ibb}.
This results in an increased sensitivity of the $l=0$ $\phi$-modes with respect to the properties of the equations of state and thus a splitting of the associated curves close to the maximum mass.

Although the differences with respect to the mean values are small, this splitting becomes clearly recognizable on the scales of the figure.
When evaluating the universal relations for these cases, this splitting together with the decreased density of points in this region leads to rather wiggly universal relations.
To avoid giving this region too much weight we have therefore decided to fit the universal relations only in the interval where there is a good agreement between the curves, as well as a high density of points, i.e., before the splitting of the curves arises (around $M/R=0.24$), as highlighted in the figure. {Meanwhile, for the general relativistic case, we fitted the data over the entire range.}

\subsection{Universal relations for the radial $\phi$-modes}

\begin{figure}[h!]
	\centering
	\resizebox{0.49\textwidth}{!}{\includegraphics{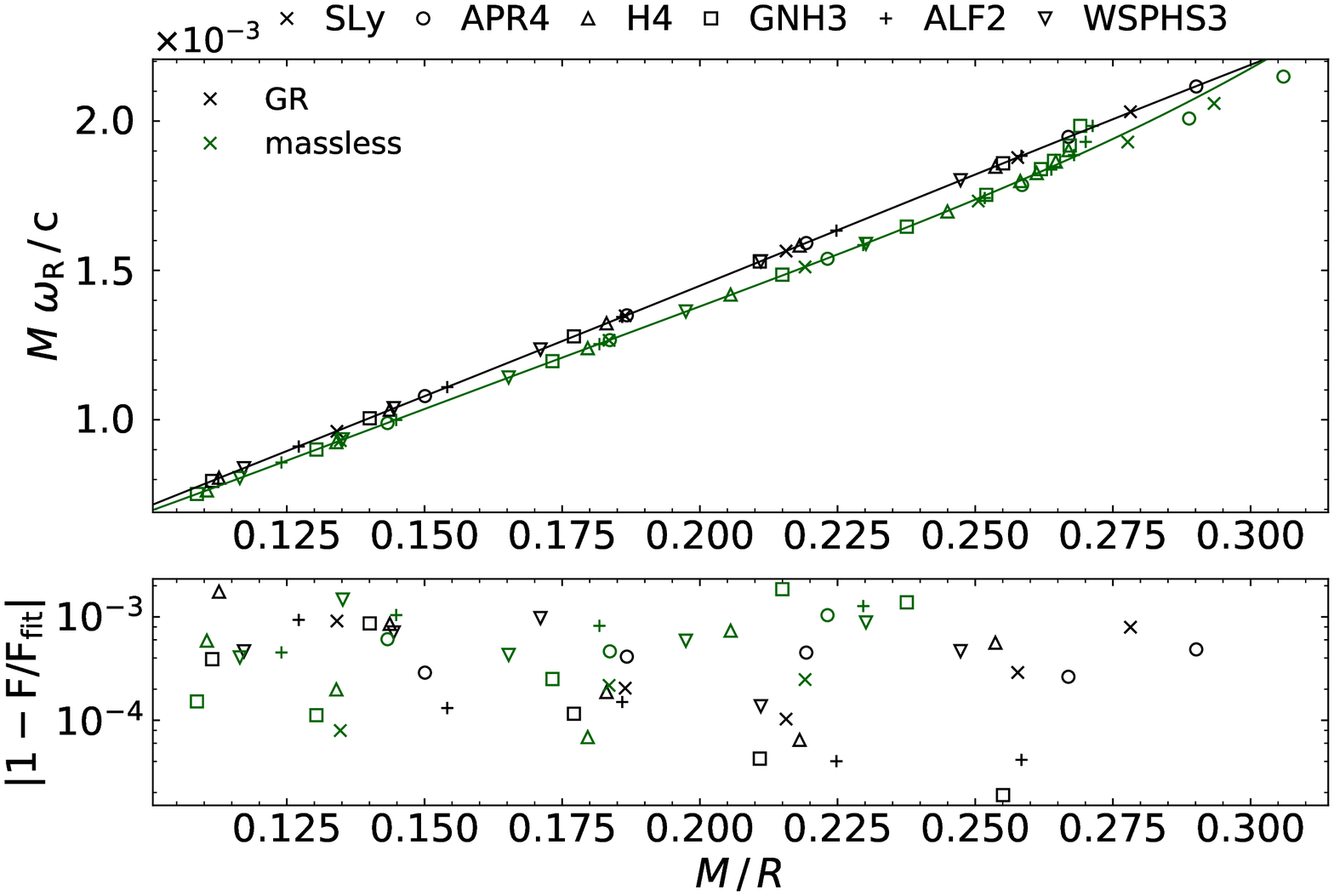}}
	\resizebox{0.49\textwidth}{!}{\includegraphics{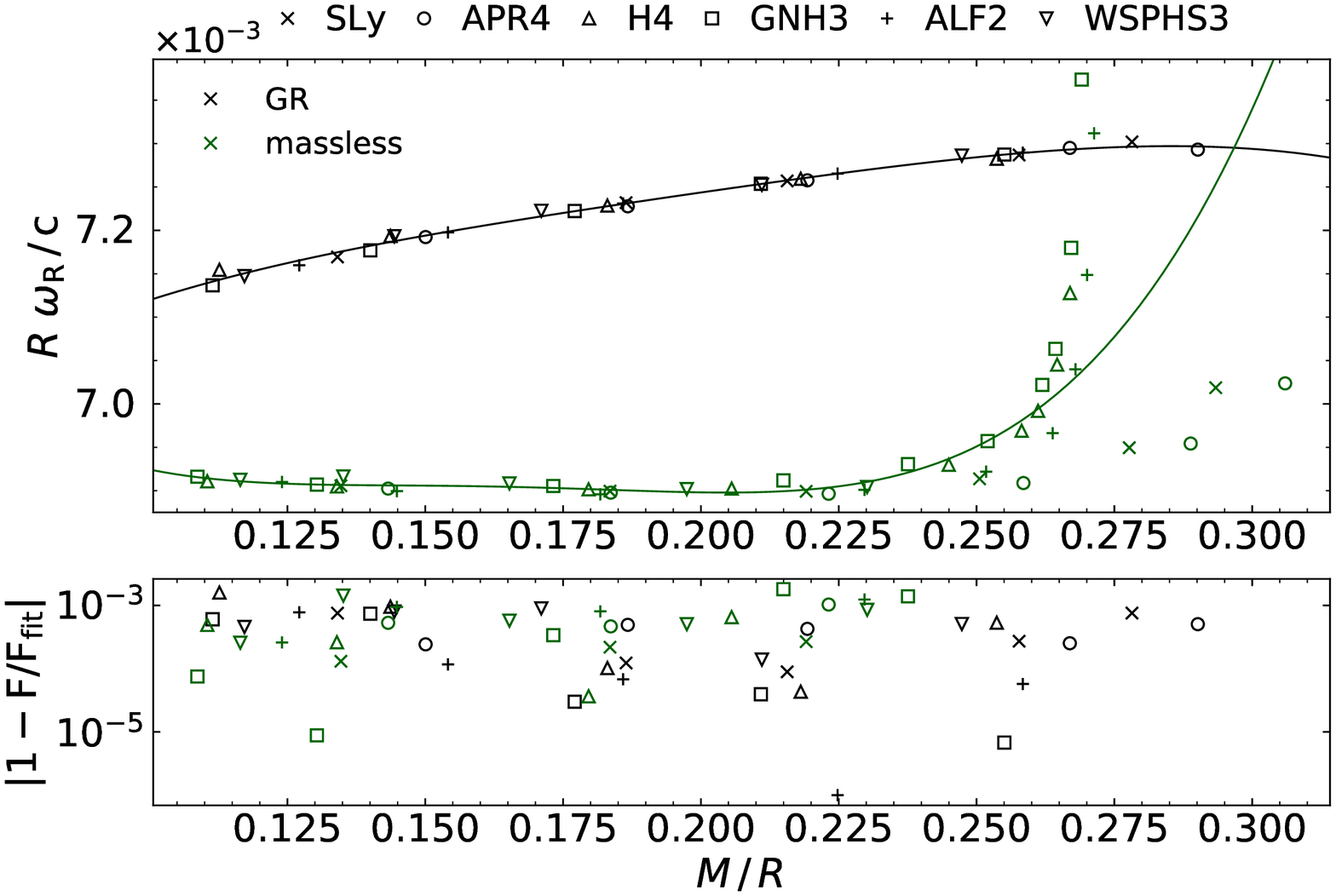}}
	\caption{Radial $\phi$-mode universal relations: dimensionless frequency $M\omega_R/c$ versus compactness $C=M/R$ (left upper panel) and fit errors (left lower panel); dimensionless frequency $R\omega_R/c$ versus compactness $C$ (right upper panel) and fit errors (right lower panel).
		The symbols indicate the respective equation of state, the massless scalar-tensor case is shown in green and the general relativistic case in black.}
	\label{fig:my_label12}
\end{figure}

Analogously to the higher modes $l$-modes we now address the universal relations for the radial $\phi$-modes.
Fig.~\ref{fig:my_label12} shows on the left the universal relations for the dimensionless frequency $M\omega_R/c$ versus the compactness $C=M/R$ and on the right for the dimensionless frequency $R\omega_R/c$ versus the compactness $C=M/R$.
In both cases the universal relations for general relativity are excellent, yielding mean errors of only 0.04\%.
The corresponding universal relations for the massless scalar-tensor theory are by far not as good.
Scaling with the mass yields a mean error of 0.9\%, 
{and scaling with the radius yields a mean error of 0.7\% when we fit over the entire range. A fit up to the compactness of $C=0.24$ yields a mean error of 0.03\%, which is comparable to the GR case.}

\begin{figure}[h!]
	\centering
	\resizebox{0.49\textwidth}{!}{\includegraphics{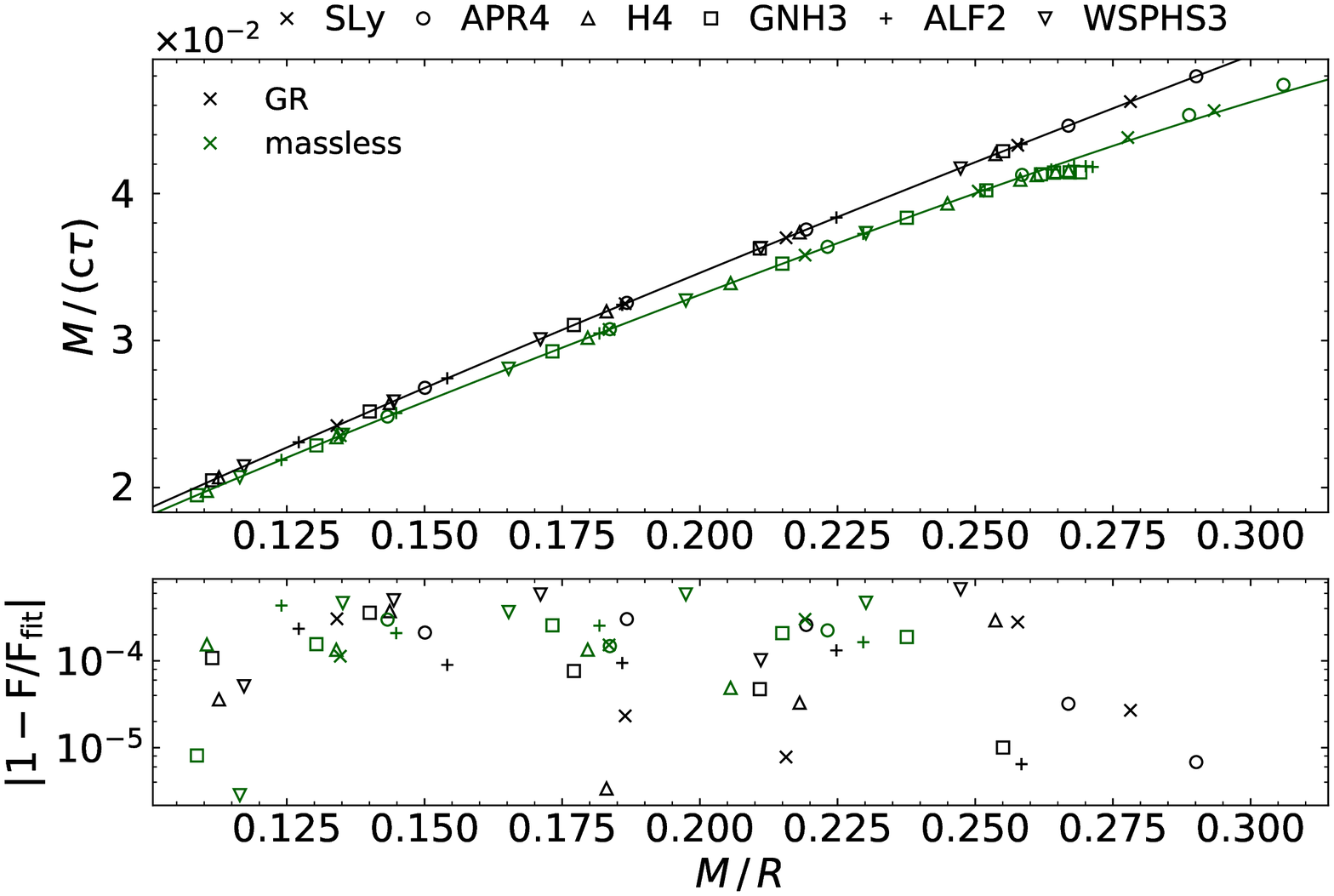}}
	\resizebox{0.49\textwidth}{!}{\includegraphics{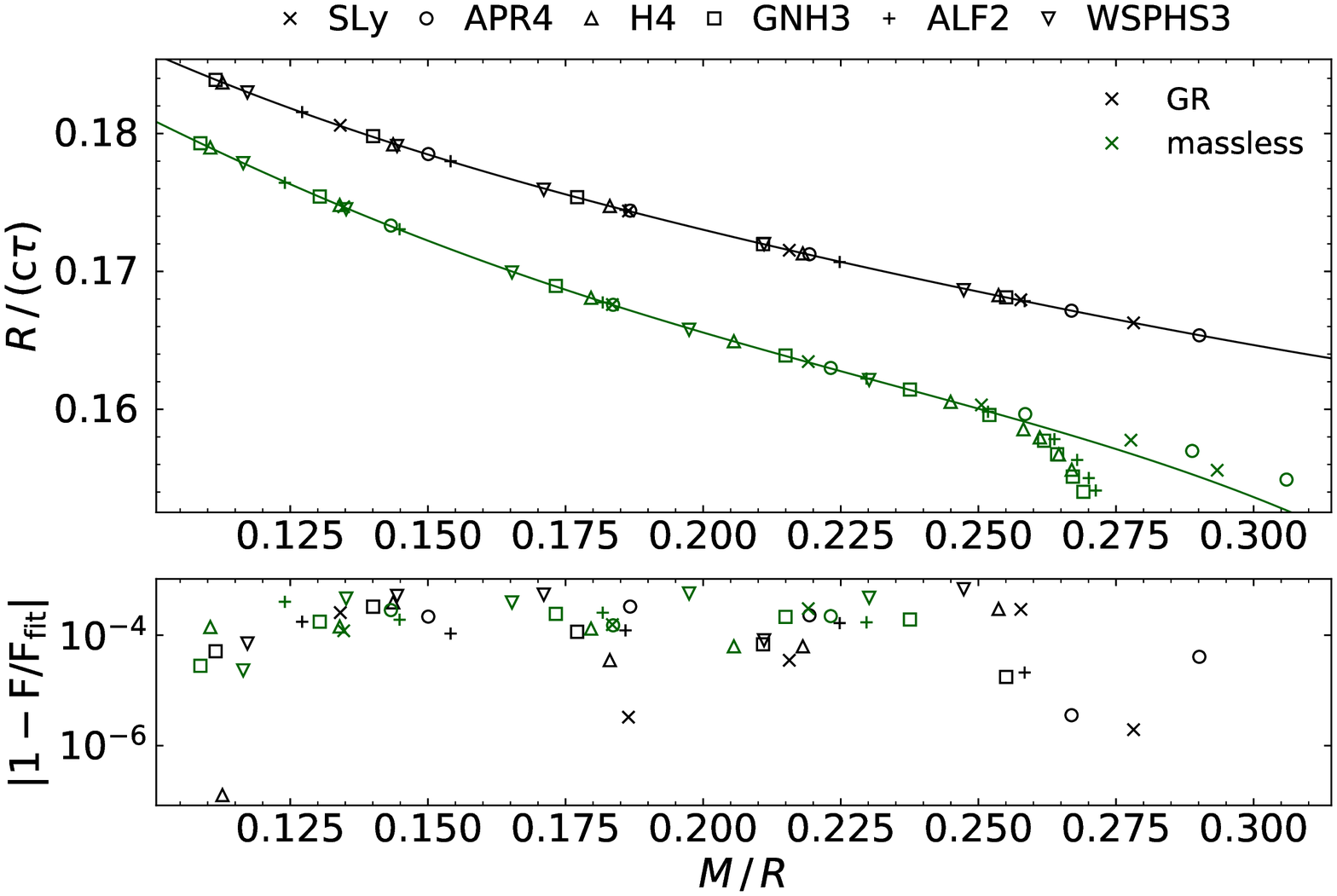}}
	\caption{Radial $\phi$-mode universal relations: dimensionless inverse damping time $M/(c\tau )$ versus compactness $C=M/R$ (left upper panel) and fit errors (left lower panel); dimensionless inverse damping time $R/(c\tau )$ versus compactness $C$ (right upper panel) and fit errors (right lower panel).
		The symbols indicate the respective equation of state, the massless scalar-tensor case is shown in green and the general relativistic case in black.}
	\label{fig:my_label13}
\end{figure}

Similar universal relations for the damping time $\tau$ are shown in Fig.~\ref{fig:my_label13}. 
Again general relativity yields excellent relations with mean errors of 0.02\%.
But here the relations of the massless scalar-tensor theory 
{
	produce} mean errors of 0.4\%. 
Here as well, a fit up to $M/R=0.24$ improves the universal relations of the massless case. The mean errors {become} 0.01\% which are of the order of the errors in GR.

\begin{figure}[h!]
	\centering
	\resizebox{0.49\textwidth}{!}{\includegraphics{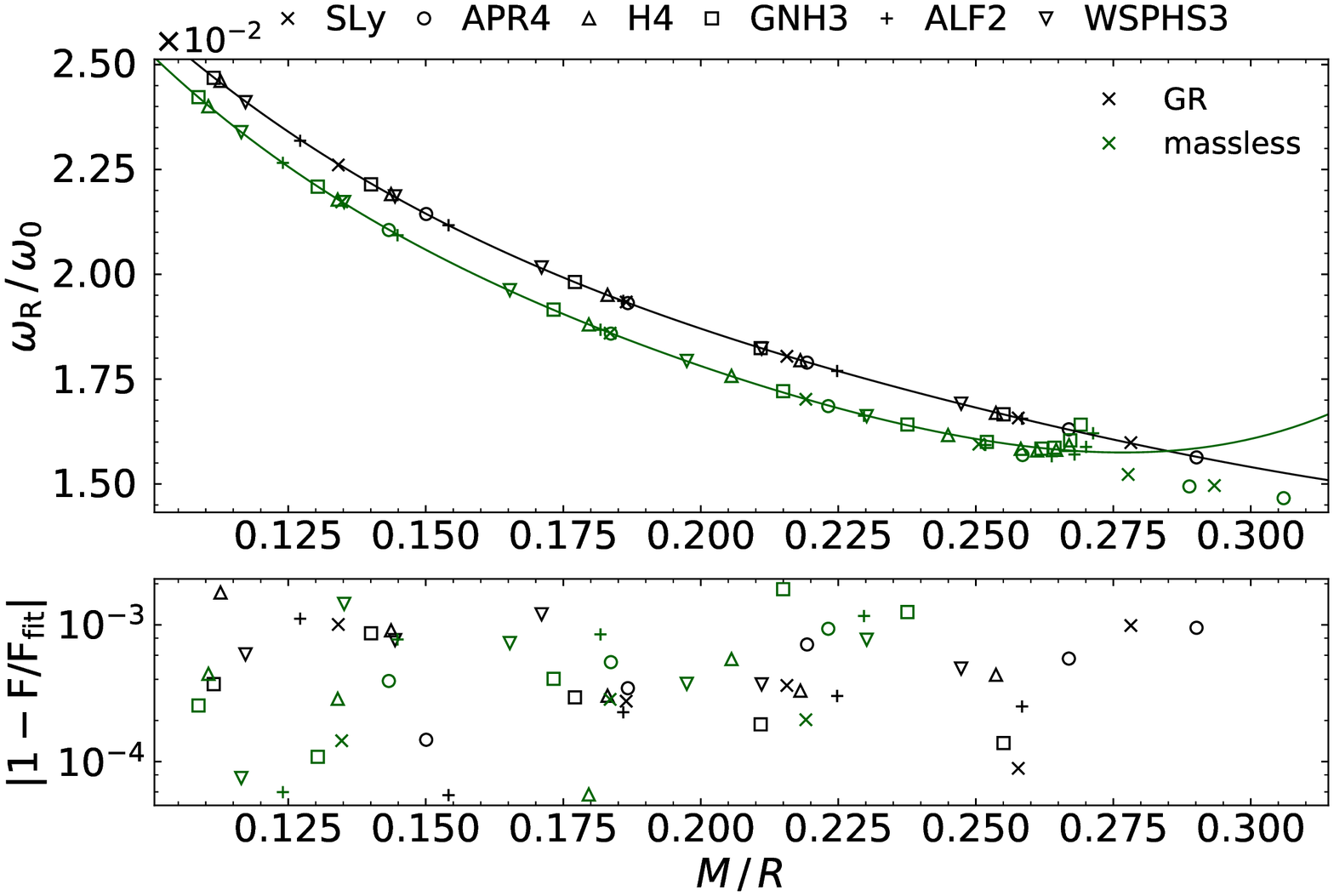}}
	\resizebox{0.49\textwidth}{!}{\includegraphics{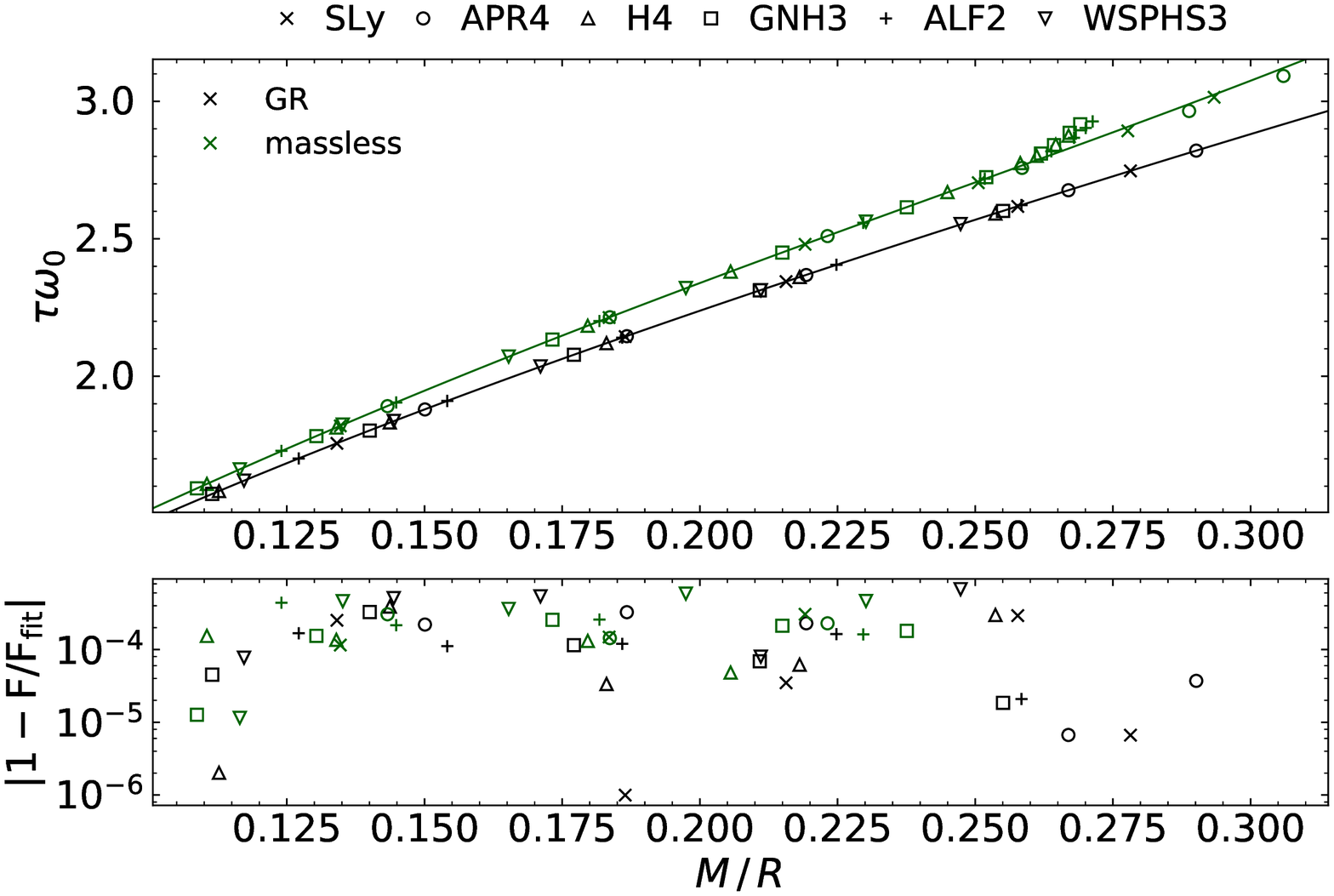}}
	\caption{Radial $\phi$-mode universal relations: dimensionless frequency $\omega_R/\omega_o$ versus compactness $C=M/R$ (left upper panel) and fit errors (left lower panel); dimensionless damping time $\tau\omega_o$ versus compactness $C$ (right upper panel) and fit errors (right lower panel).
		The symbols indicate the respective equation of state, the massless scalar-tensor case is shown in green and the general relativistic case in black.}
	\label{fig:my_label14}
\end{figure}

\begin{figure}[h!]
	\centering
	\resizebox{0.49\textwidth}{!}{\includegraphics{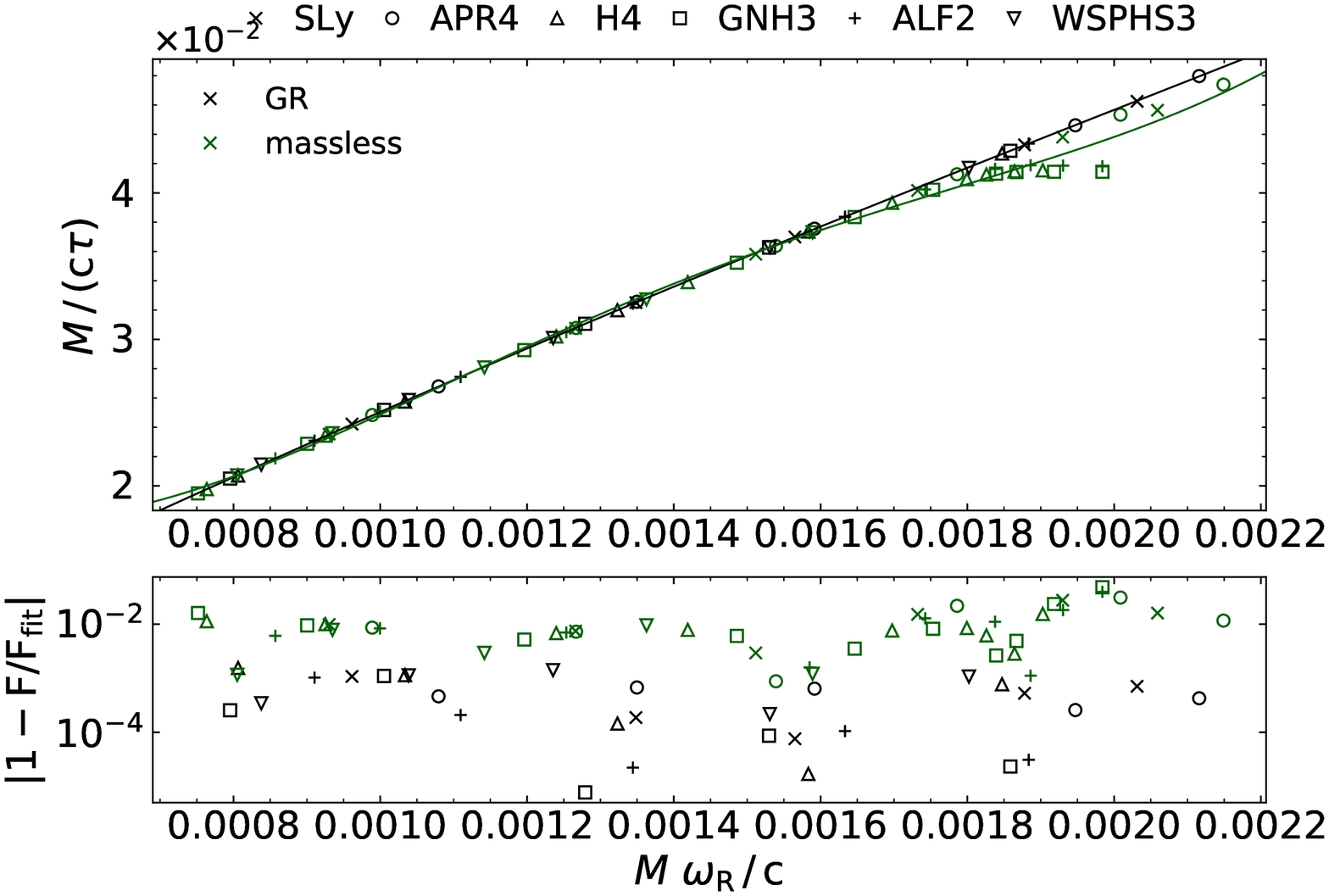}}
	\resizebox{0.49\textwidth}{!}{\includegraphics{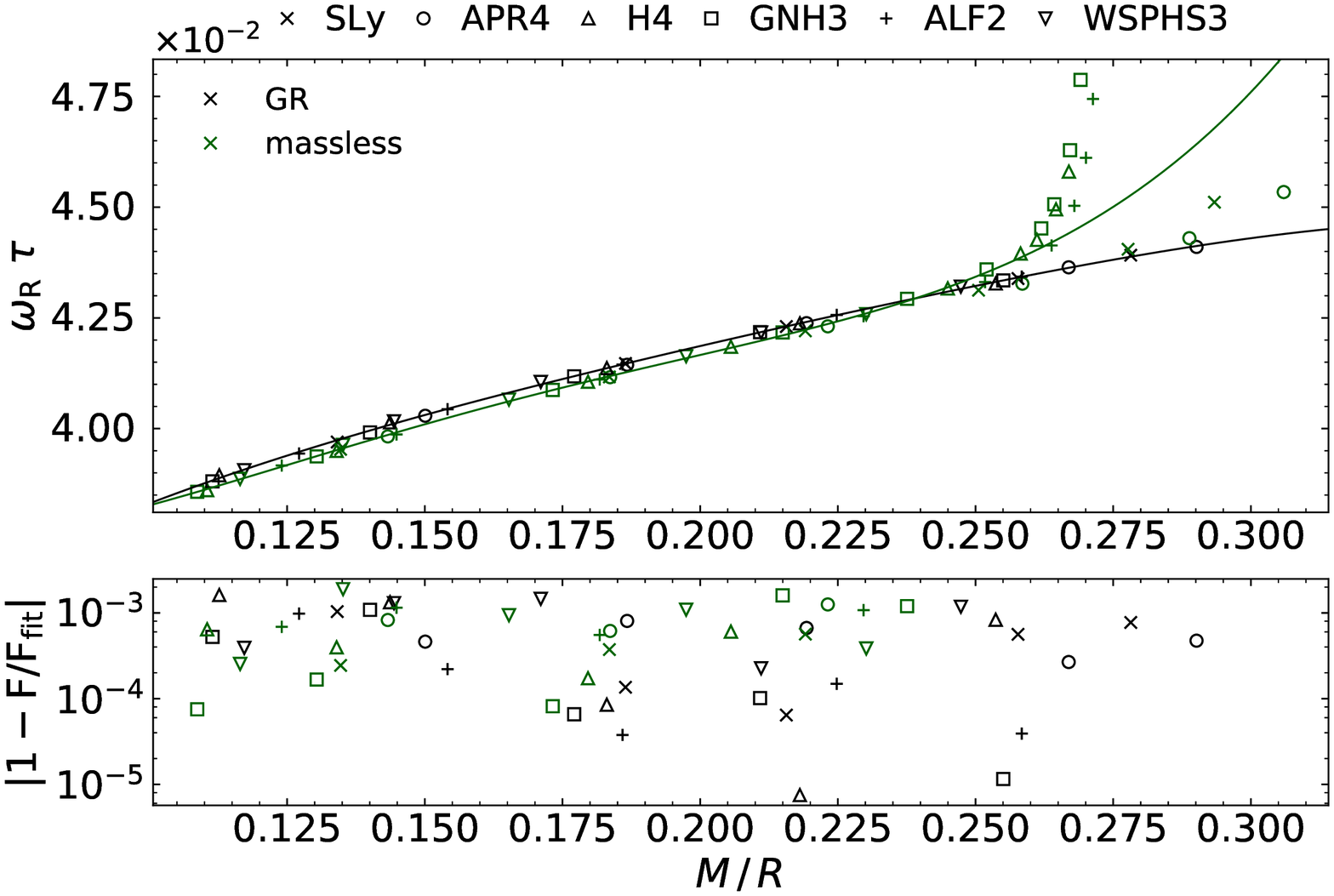}}
	\caption{Radial $\phi$-mode universal relations: dimensionless inverse damping time $M/(c\tau )$ versus dimensionless frequency $M\omega_R/c$ (left upper panel) and fit errors (left lower panel); dimensionless product $\omega_R \tau $ of frequency and damping time versus compactness $C$ (right upper panel) and fit errors (right lower panel).
		The symbols indicate the respective equation of state, the massless scalar-tensor case is shown in green and the general relativistic case in black.}
	\label{fig:my_label15}
\end{figure}

Fig.~\ref{fig:my_label14} exhibits the dimensionless frequency $\omega_R/\omega_o$ versus the compactness $C=M/R$ on the left, and the dimensionless damping time $\tau\omega_o$ versus the compactness $C$ on the right.
As before, general relativity gives excellent universal relations.
Similarly, a fit for the massless case in the {lower compactness} range provides excellent universal relations. But the discernability of the theories is not too good.
When considering the dimensionless inverse damping time $M/(c\tau )$ versus the dimensionless frequency $M\omega_R/c$ the universal relations for both theories are almost identical except for the larger masses of the stars, as seen in Fig.~\ref{fig:my_label15} (left).
This also holds for the dimensionless product $\omega_R \tau $ of the frequency and the damping time versus the compactness $C$, shown in Fig.~\ref{fig:my_label15} (right). 
As observed in all these cases, the universal relations for general relativity are excellent, whereas the universal relations for the massless scalar-tensor theory are less good, in particular, for the larger neutron star masses. 
{
	That is, the quality of the universal relations in the massless theory is comparable to the GR case only in the lower range of the compactness.}


\section{Conclusions}

Universal relations of neutron stars represent valuable tools to test the viability of alternative gravity theories, as well as, in the future, to put bounds on them with high precision gravitational wave observatories.
Here we have studied a particular set of such universal relations, that arise in a Brans-Dicke type massless scalar-tensor theory, and compared with their counterparts in general relativity.
This type of theory is obtained as a particular limit of an $f(R)$ theory, where general relativity provides the other limit, endowing the present study with theoretical interest.

The presence of a scalar degree of freedom leads to a rich spectrum of neutrons stars.
The scalar field allows for the emission of monopole and dipole radiation of the stars, that would be prohibited otherwise.
Moreover, the now propagating fluid monopole and dipole modes but also all higher multipole modes are supplemented with a new set of quasinormal modes, that are dominated by the scalar field, dubbed $\phi$-modes.
It is these $\phi$-modes on which we have focused the present study.

In order to be able to extract universal relations for the modes, and thus demonstrate (almost) independence of the equation of state employed, we have considered a set of six realistic equations of state, covering different possible star compositions, namely plain nuclear matter, nucleon-hyperon fluids, and hybrid nuclear-quark matter.
We have then tested a large variety of ways of scaling the frequency $\omega_R$ and the damping time $\tau$ to obtain dimensionless quantities (in geometric units) and considering these as functions of other dimensionless variables like the compactness or the generalized compactness.
A best fit to all the resulting points has then yielded the sought-after respective universal relations, provided the error is sufficiently small.

We have presented sets of universal relations for the quadrupole $\phi$-modes, the dipole $\phi$-modes and the radial $\phi$-modes, both in the massless scalar-tensor theory and in general relativity with a minimally coupled scalar field.
In all cases we have found very good universal relations with only small deviations from the best fits, but we have also obtained a number of rather unconvincing relations with large errors.
Interestingly, the simple scaling with the mass works mostly  quite well for these $\phi$-modes, when they are considered versus the compactness.
For the potential use of such universal relations besides the required smallness of the errors it is, however, also relevant, that the universal relations for different theories differ sufficiently as to discern them.

Having now provided the $\phi$-modes and their universal relations for the limiting theories of general relativity and the massless scalar-tensor theory, we should as our next step calculate the $\phi$-modes for finite values of the scalar field mass and extract the corresponding universal relations, as previously done for the fluid modes \citep{Blazquez-Salcedo:2020ibb,Blazquez-Salcedo:2021exm,Blazquez-Salcedo:2022pwc,Blazquez-Salcedo:2022dxh}, and the current investigations could serve as a guide in this endeavour.
Moreover, the study of the polar modes of neutron stars and their universal relations in alternative theories of gravity has just begun, and numerous interesting alternative gravities are waiting to be explored.


\newpage

\section*{Appendix 1: Tables for universal relations for quadrupole $\phi$-mode.}\label{Tables-quad}

We here present tables for the average error
of all the universal relations we tested for the quadrupole $\phi$-modes.
The average error $\bar{\epsilon}$ is given by
\begin{equation}
\bar{\epsilon} = \frac{1}{N}\sum\limits_{k=1}^N \left|1-\frac{F_k}{F_{\mathrm{fit,}k}}\right|,
\end{equation}
where $N$ is the total number of points for each theory.

\begin{table}[h!]
	\centering
	\caption{Average error $\bar{\epsilon}$ in \% for universal relations for quadrupole $\phi$-mode when plotting against compactness $C$ (left) and against generalized compactness $\eta$ (right).}
	\label{tab:average_error_l2}
	\begin{minipage}[t]{0.45\textwidth}\vspace{0pt}
		\begin{tabular}{|l||c|c|}
			\hline
			& \textbf{GR} & \textbf{massless}\\
			\hline
			$\boldsymbol{M\omega_R/c}$ & $0.1$ & $0.5$\\
			\hline
			$\boldsymbol{M/(c\tau)}$ & $1.5$ & $1.0$\\
			\hline
			$\boldsymbol{\omega_R/\omega_\mathrm{0}}$ & $0.1$ & $0.6$\\
			\hline
			$\boldsymbol{\omega_R/\hat{\omega}_\mathrm{0}}$ & $1.3$ & $2.9$\\
			\hline
			$\boldsymbol{\tau\omega_\mathrm{0}}$ & $1.6$ & $1.0$\\
			\hline
			$\boldsymbol{\tau\hat{\omega}_\mathrm{0}}$ & $2.9$ & $3.7$\\
			\hline
			$\boldsymbol{R\omega_R/c}$ & $0.1$ & $0.5$\\
			\hline
			$\boldsymbol{R\omega_R/(cC)}$ & $0.1$ & $0.6$\\
			\hline
			$\boldsymbol{R\omega_R/(cC^2)}$ & $0.6$ & $1.0$\\
			\hline
			$\boldsymbol{R\omega_R/(cC^3)}$ & $2.4$ & $3.9$\\
			\hline
			$\boldsymbol{R/(c\tau)}$ & $1.5$ & $1.0$\\
			\hline
			$\boldsymbol{R/(c\tau C)}$ & $1.5$ & $1.0$\\
			\hline
			$\boldsymbol{R/(c\tau C^2)}$ & $1.8$ & $1.6$\\
			\hline
			$\boldsymbol{R/(c\tau C^3)}$ & $4.8$ & $5.4$\\
			\hline
			$\boldsymbol{M\tau\omega_R^2/c}$ & $1.8$ & $0.3$\\
			\hline
			$\boldsymbol{R\tau\omega_R^2/c}$ & $1.7$ & $0.3$\\
			\hline
			$\boldsymbol{\omega_R\tau}$ & $1.6$ & $0.5$\\
			\hline
		\end{tabular}
	\end{minipage}
	\begin{minipage}[t]{0.45\textwidth}\vspace{0pt}
		\begin{tabular}{|l||c|c|}
			\hline
			& \textbf{GR} & \textbf{massless}\\
			\hline
			$\boldsymbol{M\omega_R/c}$ & $1.2$ & $1.7$\\
			\hline
			$\boldsymbol{M/(c\tau)}$ & $2.1$ & $1.8$\\
			\hline
			$\boldsymbol{\omega_R/\omega_\mathrm{0}}$ & $0.7$ & $0.5$\\
			\hline
			$\boldsymbol{\omega_R/\hat{\omega}_\mathrm{0}}$ & $1.1$ & $1.7$\\
			\hline
			$\boldsymbol{\tau\omega_\mathrm{0}}$ & $0.5$ & $0.5$\\
			\hline
			$\boldsymbol{\tau\hat{\omega}_\mathrm{0}}$ & $2.1$ & $1.8$\\
			\hline
			$\boldsymbol{R\omega_R/c}$ & $0.1$ & $0.4$\\
			\hline
			$\boldsymbol{R\omega_R/(cC)}$ & $1.4$ & $1.1$\\
			\hline
			$\boldsymbol{R\omega_R/(cC^2)}$ & $2.7$ & $2.7$\\
			\hline
			$\boldsymbol{R\omega_R/(cC^3)}$ & $4.7$ & $5.2$\\
			\hline
			$\boldsymbol{R/(c\tau)}$ & $0.9$ & $0.6$\\
			\hline
			$\boldsymbol{R/(c\tau C)}$ & $0.8$ & $1.1$\\
			\hline
			$\boldsymbol{R/(c\tau C^2)}$ & $2.2$ & $2.7$\\
			\hline
			$\boldsymbol{R/(c\tau C^3)}$ & $5.3$ & $6.2$\\
			\hline
			$\boldsymbol{M\tau\omega_R^2/c}$ & $0.8$ & $1.7$\\
			\hline
			$\boldsymbol{R\tau\omega_R^2/c}$ & $1.1$ & $0.4$\\
			\hline
			$\boldsymbol{\omega_R\tau}$ & $1.0$ & $0.3$\\
			\hline
			$\boldsymbol{\hat{R}\omega_R/c}$ & $1.2$ & $1.7$\\
			\hline
			$\boldsymbol{\hat{R}\omega_R/(c\eta)}$ & $1.1$ & $1.7$\\
			\hline
			$\boldsymbol{\hat{R}\omega_R/(c\eta^2)}$ & $1.1$ & $1.7$\\
			\hline
			$\boldsymbol{\hat{R}\omega_R/(c\eta^3)}$ & $1.1$ & $1.6$\\
			\hline
			$\boldsymbol{\hat{R}/(c\tau)}$ & $2.1$ & $1.8$\\
			\hline
			$\boldsymbol{\hat{R}/(c\tau\eta)}$ & $2.0$ & $1.8$\\
			\hline
			$\boldsymbol{\hat{R}/(c\tau\eta^2)}$ & $2.0$ & $1.8$\\
			\hline
			$\boldsymbol{\hat{R}/(c\tau\eta^3)}$ & $1.8$ & $1.7$\\
			\hline
		\end{tabular}
	\end{minipage}
\end{table}

\newpage
\section*{Appendix 2: Tables for universal relations for dipole $\phi$-mode.}\label{Tables-dipole}

{We here present tables for the average error of all the universal relations we tested for the dipole $\phi$-modes.}

\begin{table}[h!]
	\centering
	\caption{Average error $\bar{\epsilon}$ in \% for universal relations for dipole $\phi$-mode when plotting against compactness $C$ (left) and against generalized compactness $\eta$ (right).}
	\label{tab:average_error_l1}
	\begin{minipage}[t]{0.45\textwidth}\vspace{0pt}
		\begin{tabular}{|l||c|c|}
			\hline
			& \textbf{GR} & \textbf{massless}\\
			\hline
			$\boldsymbol{M\omega_R/c}$ & $0.4$ & $0.3$\\
			\hline
			$\boldsymbol{M/(c\tau)}$ & $1.0$ & $1.1$\\
			\hline
			$\boldsymbol{\omega_R/\omega_\mathrm{0}}$ & $0.4$ & $0.3$\\
			\hline
			$\boldsymbol{\omega_R/\hat{\omega}_\mathrm{0}}$ & $1.9$ & $1.7$\\
			\hline
			$\boldsymbol{\tau\omega_\mathrm{0}}$ & $1.0$ & $1.1$\\
			\hline
			$\boldsymbol{\tau\hat{\omega}_\mathrm{0}}$ & $1.8$ & $1.5$\\
			\hline
			$\boldsymbol{R\omega_R/c}$ & $0.4$ & $0.3$\\
			\hline
			$\boldsymbol{R\omega_R/(cC)}$ & $0.5$ & $0.3$\\
			\hline
			$\boldsymbol{R\omega_R/(cC^2)}$ & $0.9$ & $0.9$\\
			\hline
			$\boldsymbol{R\omega_R/(cC^3)}$ & $4.0$ & $4.5$\\
			\hline
			$\boldsymbol{R/(c\tau)}$ & $1.0$ & $1.1$\\
			\hline
			$\boldsymbol{R/(c\tau C)}$ & $1.0$ & $1.1$\\
			\hline
			$\boldsymbol{R/(c\tau C^2)}$ & $1.6$ & $1.7$\\
			\hline
			$\boldsymbol{R/(c\tau C^3)}$ & $5.1$ & $5.5$\\
			\hline
			$\boldsymbol{M\tau\omega_R^2/c}$ & $1.5$ & $1.5$\\
			\hline
			$\boldsymbol{R\tau\omega_R^2/c}$ & $1.5$ & $1.5$\\
			\hline
			$\boldsymbol{\omega_R\tau}$ & $1.2$ & $1.3$\\
			\hline
		\end{tabular}
	\end{minipage}
	\begin{minipage}[t]{0.45\textwidth}\vspace{0pt}
		\begin{tabular}{|l||c|c|}
			\hline
			& \textbf{GR} & \textbf{massless}\\
			\hline
			$\boldsymbol{M\omega_R/c}$ & $1.8$ & $1.6$\\
			\hline
			$\boldsymbol{M/(c\tau)}$ & $1.8$ & $1.6$\\
			\hline
			$\boldsymbol{\omega_R/\omega_\mathrm{0}}$ & $0.4$ & $0.5$\\
			\hline
			$\boldsymbol{\omega_R/\hat{\omega}_\mathrm{0}}$ & $1.8$ & $1.6$\\
			\hline
			$\boldsymbol{\tau\omega_\mathrm{0}}$ & $1.2$ & $1.3$\\
			\hline
			$\boldsymbol{\tau\hat{\omega}_\mathrm{0}}$ & $1.8$ & $1.6$\\
			\hline
			$\boldsymbol{R\omega_R/c}$ & $0.5$ & $0.4$\\
			\hline
			$\boldsymbol{R\omega_R/(cC)}$ & $1.1$ & $1.1$\\
			\hline
			$\boldsymbol{R\omega_R/(cC^2)}$ & $2.5$ & $2.6$\\
			\hline
			$\boldsymbol{R\omega_R/(cC^3)}$ & $4.3$ & $4.8$\\
			\hline
			$\boldsymbol{R/(c\tau)}$ & $0.9$ & $0.9$\\
			\hline
			$\boldsymbol{R/(c\tau C)}$ & $1.8$ & $2.0$\\
			\hline
			$\boldsymbol{R/(c\tau C^2)}$ & $3.3$ & $3.5$\\
			\hline
			$\boldsymbol{R/(c\tau C^3)}$ & $5.3$ & $5.8$\\
			\hline
			$\boldsymbol{M\tau\omega_R^2/c}$ & $2.6$ & $2.6$\\
			\hline
			$\boldsymbol{R\tau\omega_R^2/c}$ & $1.4$ & $1.4$\\
			\hline
			$\boldsymbol{\omega_R\tau}$ & $1.1$ & $1.1$\\
			\hline
			$\boldsymbol{\hat{R}\omega_R/c}$ & $1.8$ & $1.6$\\
			\hline
			$\boldsymbol{\hat{R}\omega_R/(c\eta)}$ & $1.8$ & $1.6$\\
			\hline
			$\boldsymbol{\hat{R}\omega_R/(c\eta^2)}$ & $1.8$ & $1.6$\\
			\hline
			$\boldsymbol{\hat{R}\omega_R/(c\eta^3)}$ & $1.8$ & $1.6$\\
			\hline
			$\boldsymbol{\hat{R}/(c\tau)}$ & $1.8$ & $1.6$\\
			\hline
			$\boldsymbol{\hat{R}/(c\tau\eta)}$ & $1.8$ & $1.6$\\
			\hline
			$\boldsymbol{\hat{R}/(c\tau\eta^2)}$ & $1.8$ & $1.6$\\
			\hline
			$\boldsymbol{\hat{R}/(c\tau\eta^3)}$ & $1.9$ & $1.7$\\
			\hline
		\end{tabular}
	\end{minipage}
\end{table}



\newpage
\section*{Appendix 3: Tables for universal relations for radial $\phi$-mode.}\label{Table-radial}

We here present tables for the average error of all the universal relations we tested for the radial $\phi$-modes.

\begin{table}[h!]
	\centering
	\caption{Average error $\bar{\epsilon}$ in \% for universal relations for radial $\phi$-mode when plotting against compactness $C$ (left) and against generalized compactness $\eta$ (right). {The values in brackets indicate the mean error for a fit up to $M/R=0.24$.}}
	\label{tab:average_error_l0}
	\begin{minipage}[t]{0.45\textwidth}\vspace{0pt}
		\begin{tabular}{|l||c|c|}
			\hline
			& \textbf{GR} & \textbf{massless}\\
			\hline
			$\boldsymbol{M\omega_R/c}$ & $0.04$ & $0.9\,[0.03]$\\
			\hline
			$\boldsymbol{M/(c\tau)}$ & $0.02$ & $0.4\,[0.01]$\\
			\hline
			$\boldsymbol{\omega_R/\omega_\mathrm{0}}$ & $0.05$ & $0.7\,[0.03]$\\
			\hline
			$\boldsymbol{\omega_R/\hat{\omega}_\mathrm{0}}$ & $1.5$ & $1.2\,[0.7]$\\
			\hline
			$\boldsymbol{\tau\omega_\mathrm{0}}$ & $0.02$ & $0.4\,[0.01]$\\
			\hline
			$\boldsymbol{\tau\hat{\omega}_\mathrm{0}}$ & $1.5$ & $1.7\,[0.7]$\\
			\hline
			$\boldsymbol{R\omega_R/c}$ & $0.04$ & $0.7\,[0.03]$\\
			\hline
			$\boldsymbol{R\omega_R/(cC)}$ & $0.1$ & $0.7\,[0.03]$\\
			\hline
			$\boldsymbol{R\omega_R/(cC^2)}$ & $0.6$ & $1.4\,[0.1]$\\
			\hline
			$\boldsymbol{R\omega_R/(cC^3)}$ & $2.8$ & $4.9\,[0.4]$\\
			\hline
			$\boldsymbol{R/(c\tau)}$ & $0.02$ & $0.4\,[0.01]$\\
			\hline
			$\boldsymbol{R/(c\tau C)}$ & $0.1$ & $0.3\,[0.02]$\\
			\hline
			$\boldsymbol{R/(c\tau C^2)}$ & $0.8$ & $1.2\,[0.1]$\\
			\hline
			$\boldsymbol{R/(c\tau C^3)}$ & $3.2$ & $5.6\,[0.5]$\\
			\hline
			$\boldsymbol{M\tau\omega_R^2/c}$ & $0.1$ & $2.5\,[0.07]$\\
			\hline
			$\boldsymbol{R\tau\omega_R^2/c}$ & $0.1$ & $1.9\,[0.07]$\\
			\hline
			$\boldsymbol{\omega_R\tau}$ & $0.06$ & $1.1\,[0.04]$\\
			\hline
		\end{tabular}
	\end{minipage}
	\begin{minipage}[t]{0.45\textwidth}\vspace{0pt}
		\begin{tabular}{|l||c|c|}
			\hline
			& \textbf{GR} & \textbf{massless}\\
			\hline
			$\boldsymbol{M\omega_R/c}$ & $1.4$ & $1.1$\\
			\hline
			$\boldsymbol{M/(c\tau)}$ & $1.2$ & $1.5$\\
			\hline
			$\boldsymbol{\omega_R/\omega_\mathrm{0}}$ & $0.7$ & $1.2$\\
			\hline
			$\boldsymbol{\omega_R/\hat{\omega}_\mathrm{0}}$ & $1.4$ & $1.1$\\
			\hline
			$\boldsymbol{\tau\omega_\mathrm{0}}$ & $0.8$ & $0.7$\\
			\hline
			$\boldsymbol{\tau\hat{\omega}_\mathrm{0}}$ & $1.2$ & $1.4$\\
			\hline
			$\boldsymbol{R\omega_R/c}$ & $0.06$ & $0.7$\\
			\hline
			$\boldsymbol{R\omega_R/(cC)}$ & $1.3$ & $1.8$\\
			\hline
			$\boldsymbol{R\omega_R/(cC^2)}$ & $2.8$ & $3.4$\\
			\hline
			$\boldsymbol{R\omega_R/(cC^3)}$ & $4.8$ & $6.3$\\
			\hline
			$\boldsymbol{R/(c\tau)}$ & $0.2$ & $0.3$\\
			\hline
			$\boldsymbol{R/(c\tau C)}$ & $1.5$ & $1.3$\\
			\hline
			$\boldsymbol{R/(c\tau C^2)}$ & $3.1$ & $3.0$\\
			\hline
			$\boldsymbol{R/(c\tau C^3)}$ & $5.2$ & $6.3$\\
			\hline
			$\boldsymbol{M\tau\omega_R^2/c}$ & $1.6$ & $1.9$\\
			\hline
			$\boldsymbol{R\tau\omega_R^2/c}$ & $0.3$ & $1.8$\\
			\hline
			$\boldsymbol{\omega_R\tau}$ & $0.2$ & $1.0$\\
			\hline
			$\boldsymbol{\hat{R}\omega_R/c}$ & $1.4$ & $1.1$\\
			\hline
			$\boldsymbol{\hat{R}\omega_R/(c\eta)}$ & $1.4$ & $1.1$\\
			\hline
			$\boldsymbol{\hat{R}\omega_R/(c\eta^2)}$ & $1.4$ & $1.1$\\
			\hline
			$\boldsymbol{\hat{R}\omega_R/(c\eta^3)}$ & $1.3$ & $1.2$\\
			\hline
			$\boldsymbol{\hat{R}/(c\tau)}$ & $1.2$ & $1.4$\\
			\hline
			$\boldsymbol{\hat{R}/(c\tau\eta)}$ & $1.2$ & $1.3$\\
			\hline
			$\boldsymbol{\hat{R}/(c\tau\eta^2)}$ & $1.2$ & $1.3$\\
			\hline
			$\boldsymbol{\hat{R}/(c\tau\eta^3)}$ & $1.2$ & $1.1$\\
			\hline
		\end{tabular}
	\end{minipage}
\end{table}

\section*{Conflict of Interest Statement}

The authors declare that the research was conducted in the absence of any commercial or financial relationships that could be construed as a potential conflict of interest.

\section*{Author Contributions}

All authors have contributed substantially to this paper and
agree to be accountable for the content of the work.

\section*{Acknowledgments}
{We would like to gratefully acknowledge support by the DFG Research Training Group 1620 \textit{Models of Gravity}, DFG project Ku612/18-1, FCT project PTDC/FIS-AST/3041/2020, and the COST Actions CA15117 and CA16104. FSK thanks the Department of Theoretical Physics and IPARCOS of the Complutense University of Madrid for their hospitality.
}


\section*{Data Availability Statement}
The datasets generated and analyzed can be obtained from the authors upon request.



\bibliographystyle{plainnat}

\end{document}